\documentclass[fleqn,usenatbib]{mnras}

\usepackage[T1]{fontenc}
\usepackage{ae,aecompl}
\usepackage{listings}
\usepackage{lipsum} 

\usepackage{graphicx}	% Including figure files
\usepackage{amsmath}	% Advanced maths commands
\usepackage{amssymb}	% Extra maths symbols
\usepackage{color}
\usepackage{cprotect}
\usepackage{lipsum}
\usepackage[dvipsnames]{xcolor}
\definecolor{linkcolor}{rgb}{0,0,0.25}
\hypersetup{
  colorlinks=true,        % false: boxed links; true: colored links
  linkcolor=linkcolor,    % color of internal links
  citecolor=linkcolor,    % color of links to bibliography
  filecolor=linkcolor,    % color of file links
  urlcolor=linkcolor      % color of external links
}
\usepackage{newtxtext,newtxmath}

\makeatletter
\renewcommand{\@printed}{}
\makeatother

\newcommand{\figurename}{Figure }
\newcommand{\tablename}{Table }
\newcommand{\eqnname}{Equation }
\newcommand{\secname}{Section }

\definecolor{darkgreen}{rgb}{0.0, 0.5, 0.0}

\def\ba{\begin{align}}
\def\ea{\end{align}}
\definecolor{darkblue}{rgb}{0.0, 0., 0.7}

\title[Sgr and the Milky Way's vertical waves]{Did Sgr cause the vertical waves in the solar neighbourhood?}

 \author[Bennett \& Bovy]{
Morgan Bennett\thanks{E-mail: bennett@astro.utoronto.ca}\&
Jo Bovy\\
Department of Astronomy and Astrophysics, University of Toronto, 50 St. George Street, Toronto, Ontario, M5S 3H4, Canada
}

\date{}%\date{Accepted XXX. Received YYY; in original form ZZZ}

\pubyear{2020}
\begin{document}
\label{firstpage}
\pagerange{\pageref{firstpage}--\pageref{lastpage}}
\maketitle

\begin{abstract}
The vertical distribution of stars in the solar neighbourhood is not in equilibrium but contains a wave signature in both density and velocity space originating from a perturbation. With the discovery of the phase-space spiral in \emph{Gaia} data release 2, determining the origin of this perturbation has become even more urgent. We develop and test a fast method for calculating the perturbation from a passing satellite on the vertical component of a part of a disc galaxy. This fast method allows us to test a large variety of possible perturbations to the vertical disc very quickly. We apply our method to the range of possible perturbations to the solar neighbourhood stemming from the recent passage of the Sagittarius dwarf galaxy (Sgr), varying its mass, mass profile, and present-day position within their observational uncertainties, and its orbit within different realistic models for the Milky Way's gravitational potential. We find that we are unable to reproduce the observed asymmetry in the vertical number counts and its concomitant breathing mode in velocity space for any plausible combination of Sgr and Milky-Way properties. In all cases, either the amplitude or the perturbation wavelength of the number-count asymmetry and of the oscillations in the mean vertical velocity produced by the passage of Sgr is in large disagreement with the observations from \emph{Gaia} DR2. We conclude that Sgr cannot have caused the observed oscillations in the vertical disc or the \emph{Gaia} phase-space spiral.
\end{abstract}


%%%%%%%%%%%%%%%%%%%%%%%%%%%%%%%%%%%%%%%%%%%%%%%%%%

%%%%%%%%%%%%%%%%% BODY OF PAPER %%%%%%%%%%%%%%%%%%

%comment out \usepackage{cprotect} when removing / commenting this line}

\section{Introduction}\label{sec:intro}

Mounting evidence suggests that the Milky Way disc is not in equilibrium. While equilibrium modelling can still be important to our understanding of the Galaxy, it is becoming increasingly important that we continue studying and modelling the perturbations observed in our Milky Way. Some of the most compelling evidence that the disc is not equilibrium has come from investigations into the phase-space spiral found by \citet{antoja} in the vertical component of the disc in the solar neighbourhood in \emph{Gaia} Data Release 2 (DR2; \citealt{Gaia}). The effects of the spiral can also be seen in the vertical number count density asymmetry which was first observed by \citet{widrow12} and then studied in further detail by \citet{yanny} and \citet{ME}. The trends in the velocity structure relating to this perturbation have also been studied in great detail \citep[e.g.][]{GaiaKinematics,Hunt2019,Carillo}.

The investigation into the properties of the spiral in the past few years has been thorough. \citet{BH-GALAH} considered the properties of the spiral as a function of location in the disc, metallicity, and eccentricity of the stars.  In their investigation, the phase-space spiral was prominent when the volume they considered was varied both radially and azimuthally, which they attributed to an overall corrugation of the disc. They also found that the perturbation was likely to have occurred $\sim 500$ Myr ago, but that this value was highly dependent on the chosen disc potential. In their initial paper, \citet{antoja} estimated a time since perturbation of $\sim 500$ Myr by considering test particles moving in an anharmonic oscillator potential as a model for the vertical behaviour of the disc. \citet{Laporte2019} also looked at the phase-space spiral for different groups of stars. They found that the shape of the phase-space spiral was invariant with age, which gives an upper limit on the time since the perturbation. If it had happened too far in the past, we would not see the same signal in younger stars. It also shows that all stars responded in the same way, meaning it is a property of the disc and not individual stellar groups, which is also shown in \citet{ME}. Finally, \citet{Laporte2019} also looked at the phase-space spiral at different radii and showed that it became more loosely wound and had a larger amplitude at further radii due to the change in the underlying potential.

Many possible causes of the phase-space spiral have been proposed in the literature, falling into two main schools of thought. The first is that it might be an internal perturbation. It has been proposed that resonances with the bar might have caused the observed trends in the vertical velocity, though it was ultimately ruled out as the sole cause because the resulting perturbation was too small in amplitude \citep{Monari2015}. Spiral arms have also been considered as a possible cause of the perturbation to the disc and were shown to cause some velocity structures but are also unlikely to be the sole contributor to trends seen in the mean vertical velocity. To match observations, it requires that the spiral arms have an unusually large amplitude, otherwise the amplitude of the perturbation is too small \citep{Faure2014,Quillen2018}. It has even been suggested that the coupling of the bar and spiral could be the true cause since their interaction doubles the amplitude of the perturbation compared to the case of just spiral structure \citep{Monari2016}. Finally, it was recently proposed that the buckling of the bar might have caused the oscillations in the solar neighbourhood \citep{Khoperskov2019}. However, the bar buckling likely happened a long time ago, and as of yet, there are no investigations into how recently it must have buckled to reproduce the observed perturbation.

The second prevailing theory is that the vertical perturbation to the vertical density and velocity is due to an external perturbation such as that coming from a satellite. Some have studied the possibility that dark matter subhaloes may have caused the observed fluctuations in the density and velocity, though they were not able to reproduce the vertical signal seen in the vertical disc \citep{Chequers2018}. \citet{Laporte2019} also looked into the Large Magellenic Clouds (LMC) as a possible source of the perturbation. Ultimately, they showed that not enough time has passed since the pericentre passage of the LMC for spirals to have formed using a 3-dimensional simulation of the interaction between the Milky Way and the LMC \citep{Laporte2018}.

The most popular theory is that the Sagittarius dwarf galaxy (Sgr) is the cause of the observed perturbations to the disc. This has been a large area of attention since the discovery of the phase-space spiral. \citet{Binney2018} modelled the interaction between a satellite and the disc using the impulse approximation to displace the disc in phase-space for a qualitative  comparison to the data. They found that they were able to reproduce a spiral pattern, but the nature of the assumptions meant it could not be directly compared to the observed phase-space spiral. The interaction between Sgr and the Galactic disc has also been looked at using 3-dimensional simulations which were able to produce qualitatively similar vertical perturbations to the disc \citep[e.g.][]{Purcell2011,Gomez2013,Laporte2018}. In \citet{Laporte2019}, they found that they were able to produce a spiral in a solar neighbourhood-like area of their simulation, but the tightness of the spiral was off by a factor of $\sim2$ and the amplitude was larger by a factor of approximately 1.5. They also found that to recover a spiral of a similar amplitude to the one observed, they required an initial Sgr mass of $6\times10^{10}\,\mathrm{M_\odot}$ approximately $5-6$ Gyr ago. Finally, \citet{BH-GALAH} found that the perturbation was likely due to a perturber with a mass of $5\times10^{10}\,\mathrm{M_\odot}$ which was subsequently striped of approximately half its mass. They found that a perturber with a starting mass of $10^{10}\,\mathrm{M_\odot}$ and stripped down to $5\times10^{9}\,\mathrm{M_\odot}$ was not sufficient to generate phase-space structure. However, this is surprising given that the current day mass of the Sgr progenitor is approximately $1-4\times10^8\,\mathrm{M_\odot}$, which is significantly smaller \citep{SgrModel}.

The purpose of this paper is to suggest and investigate a one-dimensional model for the interaction between a satellite and the solar neighbourhood. In \secname \ref{sec:theory}, we outline the theoretical framework on which we base the analysis in the rest of the paper. \secname \ref{sec:simple_model} considers a simple example of our model and explores how the different parameters of the model affect the analysis. The next step is comparing the model to a one-dimensional simulation, which we do in \secname \ref{sec:1DSim}, to test how the linear perturbation compares to the non-linearized calculation and to investigate the effect of self-gravity. Finally, in \secname \ref{sec:real}, we apply our model to different orbits and mass models of Sgr to see if we can reproduce the observed asymmetry. 

Throughout the paper, we use the \texttt{galpy} Python package\footnote{\url{http://github.com/jobovy/galpy}} \citep{galpy} for many of the necessary dynamical calculations. When calculating orbits and initializing the potentials in \texttt{galpy}, we use a distance to the Galactic centre of $r_\odot=8.178$ kpc unless otherwise stated \citep{Gravity}.  For the Sun's height above the mid-plane, we use $z_\odot=20.8\,\mathrm{pc}$ \citep{ME}. Finally, for the Sun's velocity in Galactocentric coordinates, we use $U,\,V,\,W= (-11.1,\,12.24,\,7.25)\,\mathrm{km\,s^{-1}}$ \citep{Vsun}. 

\section{Theoretical framework}\label{sec:theory}

\subsection{Linear Perturbation Theory}\label{sec:LPT}

Modelling the impact of a satellite on the Galactic disc is a complicated task. To simplify the problem, we make a few scientifically motivated assumptions in this paper. The first assumption is that the vertical motion of the Galactic disc can be decoupled from the radial and rotational motion of stars in the disc. This is a frequently used assumption and has been shown to be approximately true for disc stars first by \citet[\S 3.6.2,][]{binney&tremaine} and then more thoroughly by \citet{binney2011} where they found that the vertical motion was well modelled using the adiabatic approximation.

This reduces our problem from three dimensions down to one. The second assumption we make is that the perturbation is small. This allows us to consider only first order perturbations and make further simplifications to equations later in this section. The validity of this assumption is addressed in \secname \ref{sec:1DSim} where we compare the linear calculation to the non-linear case.

We compute the perturbation to the vertical phase-space distribution, modeled as a one-dimensional system, by building on the framework of \citet{kalnajs72}. We want to see how the distribution function changes after undergoing a perturbation and more precisely, how we can relate the perturbed distribution function of the final time, $t_f$ (e.g., today) to an original equilibrium distribution function. 
First, we use Liouville's theorem, which tells us that the phase-space density is conserved along orbits. This allows us to relate the distribution function today to the initial condition of the disc before it was perturbed; expressed in action-angle coordinates this is
\begin{align}
    f(J_f,\theta_f,t_f)=f'(J',\theta',t=-\infty),
\end{align}
where $J_f$ and $\theta_f$ are the action and angle coordinates at the final time $t_f$, $J'$ and $\theta'$ are the initial action and angle, and we use $t=-\infty$ to denote that we evaluate the distribution function before the start of the perturbation (in practice, this is at a finite time in the past). We can rewrite the initial action as $J'=J_f+\Delta J$ where $\Delta J$ is the perturbation to the action by a perturbing force\footnote{Physically, it is the orbit labeled by the action $J'$ that is perturbed into that labeled by $J_f$. But the definition of $\Delta J$ is motivated by the fact that $J_f$ is where we evaluate the distribution function and $J'$ is calculated and is different for every perturbation.}. Since the distribution function, $f'$ in our last step is simply the initial condition for the disc, we can choose it to be any unperturbed equilibrium distribution function, $f_0(J)$ which is only a function of the action. We can then rewrite the perturbed distribution function as:
\begin{align}
    f(J_f,\theta_f,t_f)=f_0(J')=f_0(J_f+\Delta J).
    \label{eq:df}
\end{align}
Therefore, to find the perturbed distribution function at $t_f$, we only need to calculate the change in the action due to the perturbing potential and evaluate the equilibrium distribution function at these new values.

To calculate $\Delta J$, we consider the perturbed Hamiltonian $H$, which is the sum of the equilibrium Hamiltonian $H_0$ and the perturbing potential $\Psi$. Expressed in terms of the action-angle coordinates $(J,\theta)$ of $H_0$, we have that
\begin{equation}
    H(J,\theta) = H_0(J) + \Psi(J,\theta,t)\,.
\end{equation}
This leads to the two Hamilton equations:
\begin{align}
    \frac{\partial J}{\partial t}&= -\frac{\partial}{\partial \theta}\left(H_0+\Psi\right)= -\frac{\partial}{\partial \theta}\Psi(J,\theta,t) \label{eq:ham}\\
    \frac{\partial \theta}{\partial t}&= \,\,\,\frac{\partial}{\partial J}\left(H_0+\Psi\right)= \Omega(J)+\frac{\partial}{\partial J}\Psi(J,\theta,t)\,\label{eq:ham2}
\end{align}
where $\Omega(J)$ is the frequency of orbit $J$ in $H_0$, $\Omega(J) = \partial H_0 / \partial J$. To calculate $\Delta J$, we need to integrate \eqnname (\ref{eq:ham}) over the orbit of the disc particles from $t=t_f$ to $t=-\infty$. At the beginning of this section, we discussed our assumption that the perturbation is small. This assumption allows us to integrate \eqnname (\ref{eq:ham}) over the unperturbed orbit of a particle in the disc as opposed to the perturbed one:
\begin{align}
    \Delta J = J'-J_f = -\int_{-\infty}^{t_f}\frac{\partial}{\partial \theta}\Psi(J_f, \,\theta_f+\Omega(t_f-\Tilde{t}),\,\Tilde{t})\,d\Tilde{t}
    \label{eq:dJ}
\end{align}
It is possible to further simplify the integrand of the above equation. Using the chain rule, we can rewrite the derivative of the potential to a more familiar form.
\begin{align}
    \frac{\partial\Psi}{\partial\theta}= \frac{\partial z}{\partial\theta}\frac{\partial\Psi}{\partial z}= \frac{\partial z}{\partial t}\frac{\partial t}{\partial\theta}\frac{\partial\Psi}{\partial z} = \frac{v_z}{\Omega}\frac{\partial\Psi}{\partial z}
    \label{eq:integrand}
\end{align}
where $\Omega$ is the orbital frequency, $v_z$ is the vertical velocity, and the derivative of the potential with respect to $z$ is just minus the vertical force due to the perturber. Thus, to compute the perturbed action, we simply integrate the perturbing force, multiplied by the ratio of the vertical velocity to the vertical frequency, over the unperturbed orbit. 
\begin{align}
    \Delta J = \int_{-\infty}^{t_f}\frac{v_z}{\Omega_f}F_z(J_f, \theta_f+\Omega_f(t_f-\Tilde{t}),\Tilde{t})\,d\Tilde{t}
    \label{eq:deltaJ}
\end{align}
where $F_z$ is the vertical force from the perturbation and $v_z$ is the vertical velocity along the orbit, and therefore also a function of time. Finally, we can use the unperturbed Hamiltonian, $H_0(J)$ for the canonical transformation between action-angle coordinates and Cartesian coordinates. 
\begin{align}
    f(x,v,t)= f(J,\theta,t).
\end{align}
This allows us to compare our model to existing measurements of the perturbed disc, including the number count asymmetry \citep[e.g.][]{widrow12,yanny,ME} and the phase-space spiral \citep[e.g.][]{antoja}.

\subsection{Model Ingredients}\label{sec:ingr}

There are several factors that contribute to the calculation of the perturbed distribution function for a single point in phase-space. We need the unperturbed Hamiltonian, $H_0$, the perturbation to the action, $\Delta J$, and the equilibrium distribution function, $f_0(J)$. The unperturbed Hamiltonian allows us to convert the $(z,v_z)$ phase-space to action-angle coordinates such that we can use the framework set out above. 

To calculate the perturbation to the action given by \eqnname (\ref{eq:deltaJ}) there are two things we need to do: (i) we need the unperturbed orbit of a particle in the disc and (ii) we need the perturbing vertical force on the particle throughout its orbit. To get the unperturbed orbit of the particle, we simply integrate the orbit numerically backwards from the final position at $t_f$. The orbit gives the position and velocity at which to evaluate the perturbing force in \eqnname (\ref{eq:deltaJ}) and we compute the action $J_f$ and frequency $\Omega(J_f)$ using the usual one-dimensional integrals, implemented in \texttt{galpy}. The vertical force in equation \eqnname (\ref{eq:deltaJ}) can be any arbitrary force. In this paper, we consider the vertical force due to the passage of a satellite. This can be computed by taking the vertical component of the 3-dimensional force a satellite exerts on the solar neighbourhood as it orbits the Milky Way. This requires knowing the orbit of the satellite as well as the mass distribution of the satellite. We then calculate the force of the satellite potential on the solar neighbourhood given the location of the satellite, the location of the solar neighbourhood as it orbits the galactic centre, and the vertical location of our particle in the disc to find the relative distance between the two. The equation for this calculation is:
\begin{equation}
    F_z= F\cdot\frac{z_\mathrm{disc}-z_\mathrm{sat}}{\left|\Vec{x}_\mathrm{disc}-\Vec{x}_\mathrm{sat}\right|} = -\nabla\psi\cdot\frac{z_\mathrm{disc}-z_\mathrm{sat}}{\left|\Vec{x}_\mathrm{disc}-\Vec{x}_\mathrm{sat}\right|}\label{eq:Fz}
\end{equation}
Where $z_\mathrm{disc}$ and $z_\mathrm{sat}$ are the vertical positions of the disc and satellite respectively. The $x$ and $y$ coordinates of $\Vec{x}_\mathrm{disc}=(x_{SN},y_{SN},z_\mathrm{disc})$ is the Galactocentric position of the solar neighbourhood as it orbits the Galactic centre and $\vec{x}_\mathrm{sat}$ is the Galactocentric position of the satellite. Finally, $\psi$ is the gravitational potential of the satellite. We assume that the solar neighbourhood orbits the centre of the Milky Way on a circular orbit with a frequency of $v_c(R_0)/R_0$, where $R_0$ is the distance from the sun to the Galactic centre and $v_c(R_0)$ is the circular velocity at the solar radius.

Once we have the perturbation to the action, the second component of our model is the equilibrium distribution function. This can be any distribution function which is a function only of the action. We consider equilibrium distribution functions that are a single quasi-isothermal distribution function \citep{binney2011} or a combination of multiple such distributions, with the form
\begin{align}
    f_0(z,v)= \frac{n_0}{\sqrt{2\pi}\sigma}\exp\left[\frac{-J(z,v)\,\upnu}{\sigma^2}\right]
    \label{eq:qi_df}, 
\end{align}
where $n_0$ is the number density, $\sigma$ is a velocity-dispersion parameter, $J$ is the action of a point in phase-space, and $\upnu$ is the vertical frequency of the disc. Such distribution functions or their combination represent the stellar distribution in the solar neighbourhood well \citep{binney10,Ting,BovyRix}.

\section{Considering a simple model}\label{sec:simple_model}

\subsection{Model setup}\label{sec:model_setup}

We next use the formalism introduced above to compute the asymmetry and mean vertical velocity for a specific disc model and a given perturbing force on that disc. To better understand the dynamics of the perturbation and the effect of the various ingredients in our model, we investigate the effect of a perturbing force on the vertical phase-space of a stellar disc with a simple setup in the next few subsections. For the first component of our model, the unperturbed Hamiltonian, we use that of an isothermal disc with density
\begin{equation}
    \rho(z) = \rho_0\,\mathrm{sech}^2\left(\frac{x}{2H}\right)\,,
\end{equation}
where $H^2 = \sigma^2/[8\pi G \rho_0]$. We chose a mid-plane density of $\rho_0=0.1\,\mathrm{M_\odot\,pc^{-3}}$ and a velocity dispersion of $\sigma=20.5\,\mathrm{km\,s^{-1}}$. The mid-plane density is chosen to approximately match the observed mid-plane density of the solar neighbourhood \citep[e.g.][]{p0-1,p0-2}. We choose the velocity dispersion by finding the standard deviation of the \emph{Gaia} radial velocity sample discussed in \citet{ME}. 

\eqnname (\ref{eq:deltaJ}) requires that we integrate the orbit of the particles in the disc as well as the satellite backwards in time until before the perturbation starts. In practice, we integrate back to an apocentre. We choose to start at the furthest point in the orbit, because that is where the force is smallest and thus our assumption that the disc is originally in equilibrium is best satisfied there. We can then integrate the orbit our satellite as well as the disc particle from the apocentre time $t_a$ to our time of interest, $t_f$ and we can then repeat this calculation for a series of $t_f$s to recover the time dependence of the perturbation.

\begin{figure}
    \centering
    \includegraphics[width=0.45\textwidth]{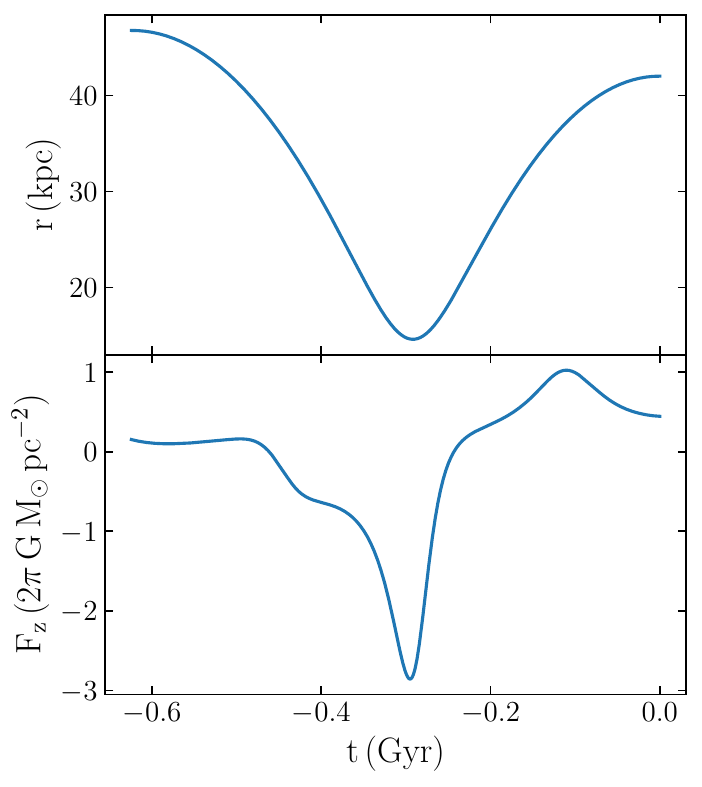}
    \caption{\textit{Top:} Galactocentric radius of our satellite with time. \textit{Bottom:} Vertical component of the force on the $(z,v_z)=(0,0)$ point in phase-space over time.}
    \label{fig:r_F}
\end{figure}

To calculate the force of a Sagittarius-like satellite on the disc, we integrate the orbit of the satellite from the apocentre to apocentre. We choose $t_f$ to be the next apocentre so that we are considering a full orbit of the satellite. \figurename \ref{fig:r_F} shows an example orbit as well as the resultant vertical force from that satellite. As mentioned in \secname \ref{sec:ingr}, we use the 3-dimensional force from the satellite potential and take the vertical component. The overall shape of the force seen in \figurename \ref{fig:r_F} is given by the orbit of the satellite. The additional smaller scale oscillations in the vertical force are due to the orbit of the solar neighbourhood around the galactic centre, which is orbiting with a frequency of $v_\odot/r_\odot=220\,\mathrm{km\,s^{-1}}/8.178\,\mathrm{kpc}=26.9\,\mathrm{km\,s^{-1}\,kpc^{-1}}$. In \secname \ref{sec:sgr_orbit}, we perform a thorough investigation of Sagittarius' orbit and we choose a sample orbit for use here from among the larger number of orbits considered there. While we explore several different Sgr potentials in \secname \ref{sec:sgr_model}, for our example here, we choose the second Sgr model, Sgr2, which consists of a stellar component represented by a Hernquist potential \citep{Hernquist} with a mass of $2\times10^8$ M$_\odot$ and a scale radius of 0.65 kpc. We also include a dark matter component as a Hernquist potential with a mass of $10^{10}$ M$_\odot$ and a scale radius of 3 kpc. These choice were made to reflect the results of \citet{SgrModel} which investigated the properties of the Sgr progenitor today. We can then use this potential and the position of the satellite at each time to calculate its force on each point in phase-space. 

These components are all we need to calculate the perturbation to the action. As discussed in \secname \ref{sec:ingr}, the final key to calculating the perturbed distribution function, is the initial unperturbed distribution function, $f_0$. For our analysis of a simple disc model, we choose to use a single quasi-isothermal distribution function. We choose this distribution function and its parameters to be consistent with the potential in which the orbits are calculated and therefore pick $n_0=0.1\,\mathrm{M_\odot\,pc^{-3}}$ and $\sigma=20.5\,\mathrm{km\,s^{-1}}$ to match the potential.

\begin{figure}
    \centering
    \includegraphics[width=0.45\textwidth]{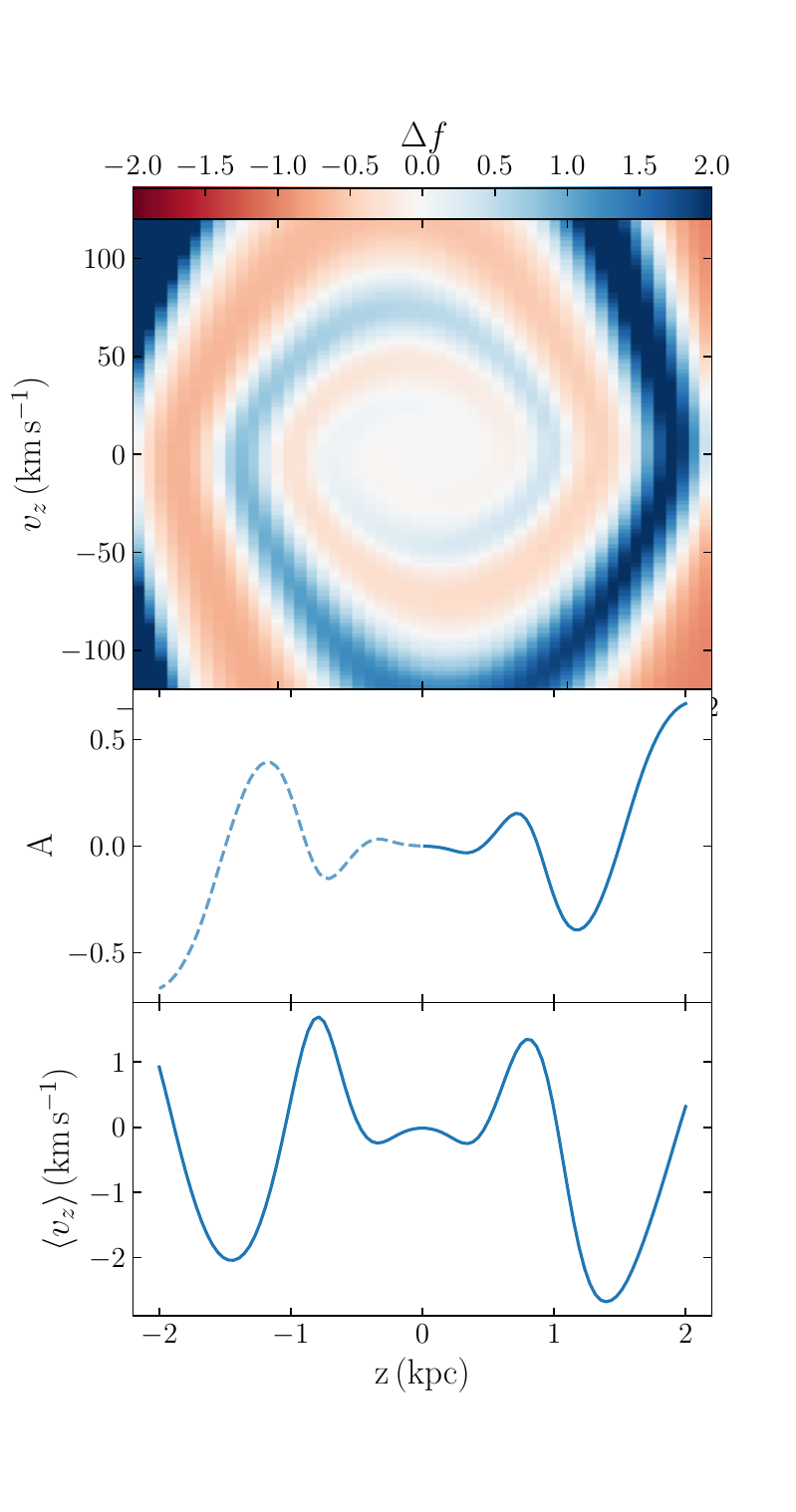}
    \caption{\textit{Top:} Spiral in the phase-space difference given in \eqnname (\ref{eq:deltaf}) as a result of the satellite perturbation at t=0 (now). \textit{Middle:} The calculated asymmetry as a function of height above the mid-plane for the example satellite perturbation on the disc. Note that the calculation of asymmetry assures it is anti-symmetric. So we plot it over $-2$ to 2 kpc, but all information is included in 0 to 2 kpc. \textit{Bottom:} Mean velocity as a function of height above and below the mid-plane.}
    \label{fig:A_vz}
\end{figure}

Using this equilibrium distribution function and the perturbed actions at a given time for each point in the $z-v_z$ phase-space, we can calculate the perturbed distribution function. We use this to recover both the perturbed density and the perturbed mean vertical velocity as a function of time. Using the perturbed number density at the final time, $n(z;t_f)$, we can calculate the vertical number count asymmetry using:
 \begin{equation}
     A=\frac{n(z;t_f)-n(-z;t_f)}{n(z;t_f)+n(-z;t_f)}\,.
 \end{equation}
\figurename \ref{fig:A_vz} shows the calculated phase-space difference, asymmetry, and mean vertical velocity for the simple model. The phase-space difference is given by:
\begin{equation}
    \Delta f= \frac{ f_0(J+\Delta J)-f_0(J)}{f_0(J)}
    \label{eq:deltaf}
\end{equation}
The left-hand side of the asymmetry in the top panel of the figure is plotted with a dashed line since the definition of the asymmetry is inherently anti-symmetric. This means that all the information is contained in the asymmetry for $z\geq0$. We choose to plot it for values of $z$ less than zero here to allow for a comparison to the phase-space spiral and trends in the mean vertical velocity. Throughout the paper, we will reference the perturbation wavelength which is a general indicator of the tightness of the phase-space spiral and therefore the wavelength of both the asymmetry and the mean vertical velocity. It is clear to see how the wavelength of the mean vertical velocity relates to the phase-space spiral because it is the first moment of the distribution function over $v_z$. The relation between the asymmetry and phase-space is somewhat more complicated because they are not directly related, but the asymmetry is a function of the zeroth moment, which becomes evident as we compare general features of phase-space to the asymmetry.

\begin{figure}
    \centering
    \includegraphics[width=0.45\textwidth]{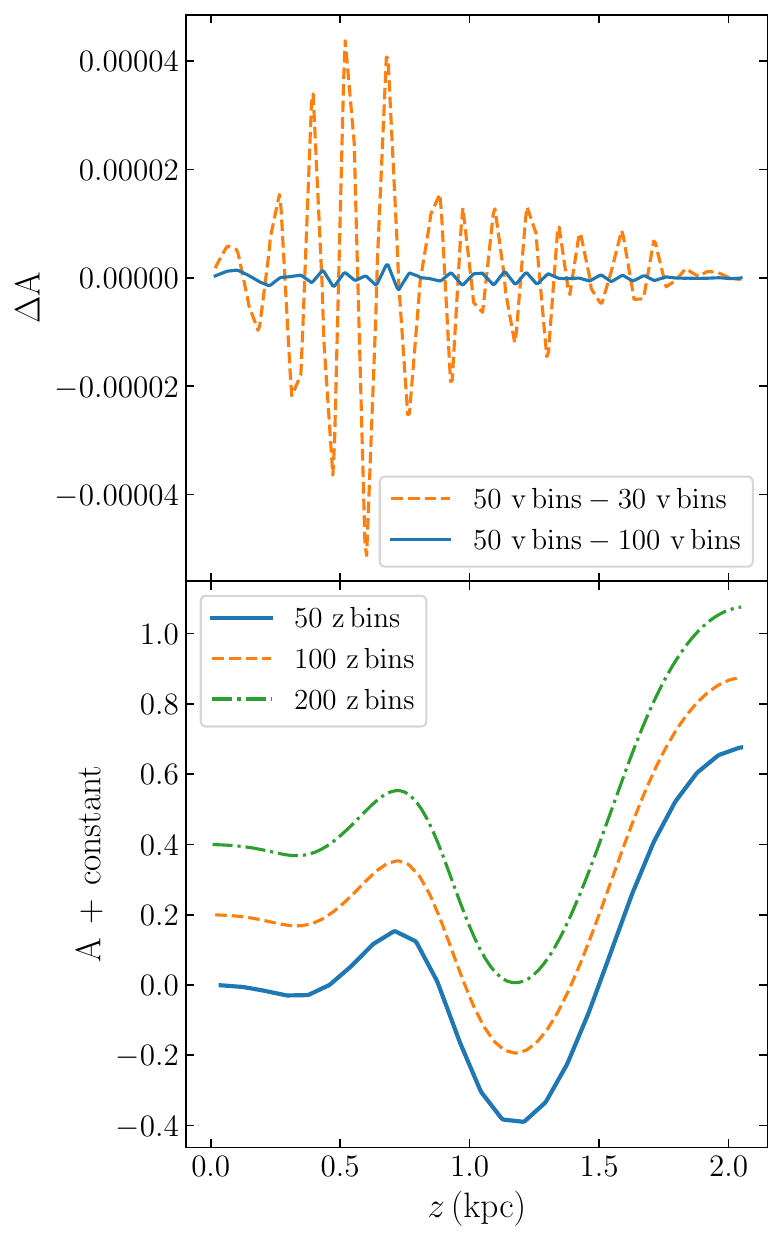}
    \caption{\textit{Top:} The absolute difference between the asymmetry when calculated using different numbers of velocity bins. The dashed orange line shows the difference between 30 and 50 bins in velocity and the solid blue line shows the difference between 50 and 100 velocity bins. \textit{Bottom:} Asymmetry for different numbers of bins in $z$. The three lines are 50 bins in $z$ (solid blue), 100 bins in $z$ (orange dashed), and 200 bins in $z$ (green dash-dotted). The asymmetry only has half the number of bins listed above.}
    \label{fig:grid_diff}
\end{figure}

With regards to the grid in phase-space, we chose to consider a range in $z$ from $-2$ kpc to $2$ kpc, sampled by 100 bins and a range in vertical velocity, $v_z$ between $-120$ km s$^{-1}$ and $120$ km s$^{-1}$ sampled by 50 bins. In choosing the grid-size in phase-space, we explored how it affects the density as you increase or decrease the number of sampling points in each dimension. If you increase the number of bins in phase-space along the $z$ direction, the values at each point are the same, and the only difference is the resolution with which you can see the asymmetry. The top panel of \figurename \ref{fig:grid_diff} shows the difference between the asymmetry using 50 and 30 velocity bins or 50 and 100 velocity bins. In both cases, the grid size along the $z$-axis is 50 bins. The largest difference between 30 and 50 velocity bins is $0.5\%$ the size of the asymmetry. However, the difference between 50 and 100 bins in velocity is always smaller than $0.07\%$ of the asymmetry. We choose to use 50 bins because as the asymmetry becomes more complicated, with more oscillations, the importance of the number of velocity bins increases. This is not surprising because with only 30 bins in velocity, each grid point is 8 km s$^{-1}$ apart which is approximately 4 times larger than the perturbations we see in the mean velocity. Therefore, we use 50 bins since it is a good balance between accuracy and computational power required.

The bottom panel of \figurename \ref{fig:grid_diff} shows the asymmetry in three different $z$ binning scenarios. In all three cases the grid size along the velocity axis is 50 bins. Since they all have the same grid spacing in velocity, the actual value of the asymmetry stays the same at each point. When calculating the asymmetry, we only have half the points as our number of bins in z, because the information at positive and negative values is redundant. The difference between the grid sizes in $z$ is the sampling of the asymmetry. With only 50 bins in z, the asymmetry looks slightly choppy. However, there is no big difference between 100 and 200 bins in z. So again, for computational reasons, we choose to use 100 grid bins in z. Again, like with the velocity bins, as the complexity of the asymmetry increases, the number of bins along $z$ and therefore the sampling, becomes even more important.

\subsection{Exploring Changes in the Mid-plane Density of the Disc}\label{sec:mpd}

\begin{figure}
    \centering
    \includegraphics[width=0.45\textwidth]{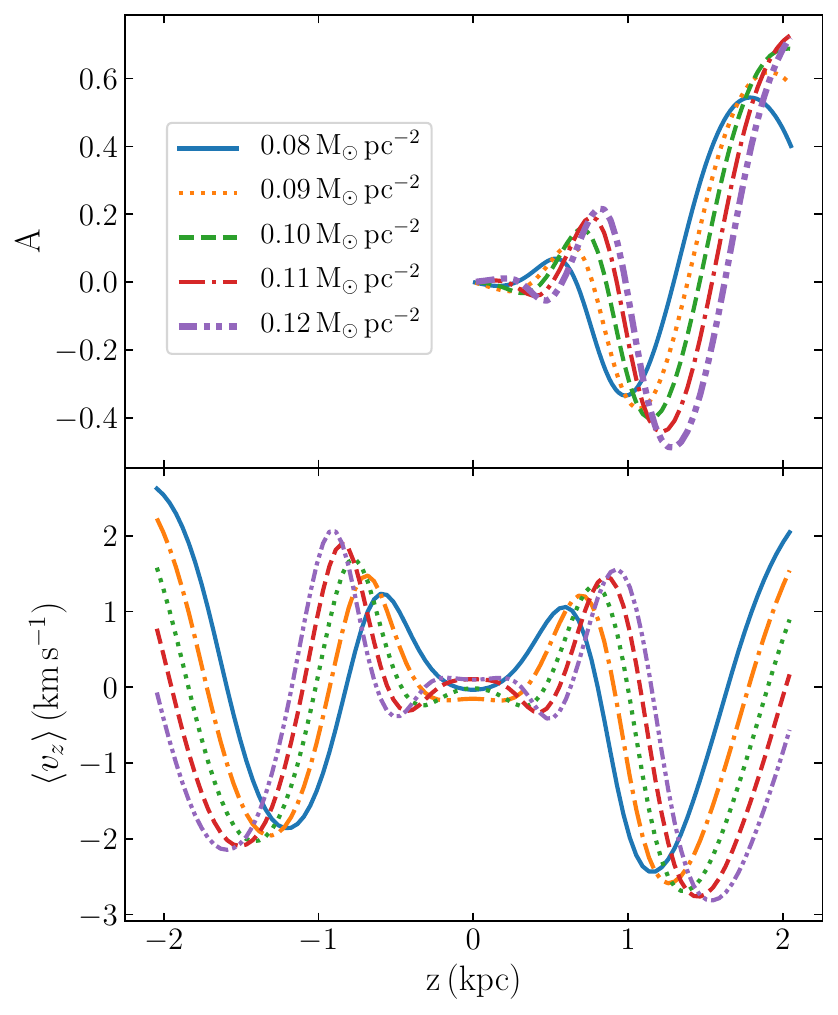}
    \caption{Effect of the disc's mid-plane density on the perturbation from the example satellite perturbation. \textit{Top:} Asymmetry for five different isothermal discs set to have the same velocity dispersion and varying mid-plane densities at $t=0$. \textit{Bottom:} Mean velocity as a function of height for the same set of models. The mid-plane density sets the vertical frequency and therefore, changing the mid-plane density changes the perturbation wavelength.}
    \label{fig:MD_pert}
\end{figure}

When investigating the effects of the disc on the asymmetry, there are two components to consider. The first is the gravitational potential we use to integrate the orbits of the disc and the second is the equilibrium distribution function we use to calculate the perturbed distribution function. To investigate how the mid-plane density affects the asymmetry, we need to make sure these two are consistent with each other. To fully investigate this, we consider five different quasi-isothermal discs with a velocity dispersion of $\sigma=20.5$ $\mathrm{km\,s^{-1}}$ and mid-plane densities of $\rho_o=$ 0.08, 0.09, 0.10, 0.11, and 0.12 $\mathrm{M_\odot pc^{-3}}$. In each case, we initialise a new disc and reintegrate the orbits for each disc to reflect the different mid-plane density. One important note is that by fixing the velocity dispersion of the discs and changing the mid-plane density, we are also changing the mass of the disc. By changing the mass of the disc, the vertical frequency of the disc is also changed.

The top panel in \figurename \ref{fig:MD_pert} shows the calculated asymmetry for the five different cases. The perturbation to the disc in each case is due to the model, orbit and satellite potential discussed in \secname \ref{sec:model_setup} and calculated at $t_f=0$. As we increase the mid-plane density and therefore the mass of the disc and the vertical frequency of the disc, we increase the both the amplitude and the frequency of the asymmetry. Of the two effects, the change to the perturbation wavelength is the more dominant change. As seen in the bottom panel of \figurename \ref{fig:MD_pert}, changing the mid-plane density has a similar effect on the mean velocity as a function of height. It increases the amplitude of the signal in the mean velocity, but also has a larger effect on the perturbation wavelength of the signal, decreasing the perturbation wavelength as we increase the mid-plane density.

\subsection{Exploring the difference of tracers in the disc}\label{sec:vd}

\begin{figure}
    \centering
    \includegraphics[width=0.45\textwidth]{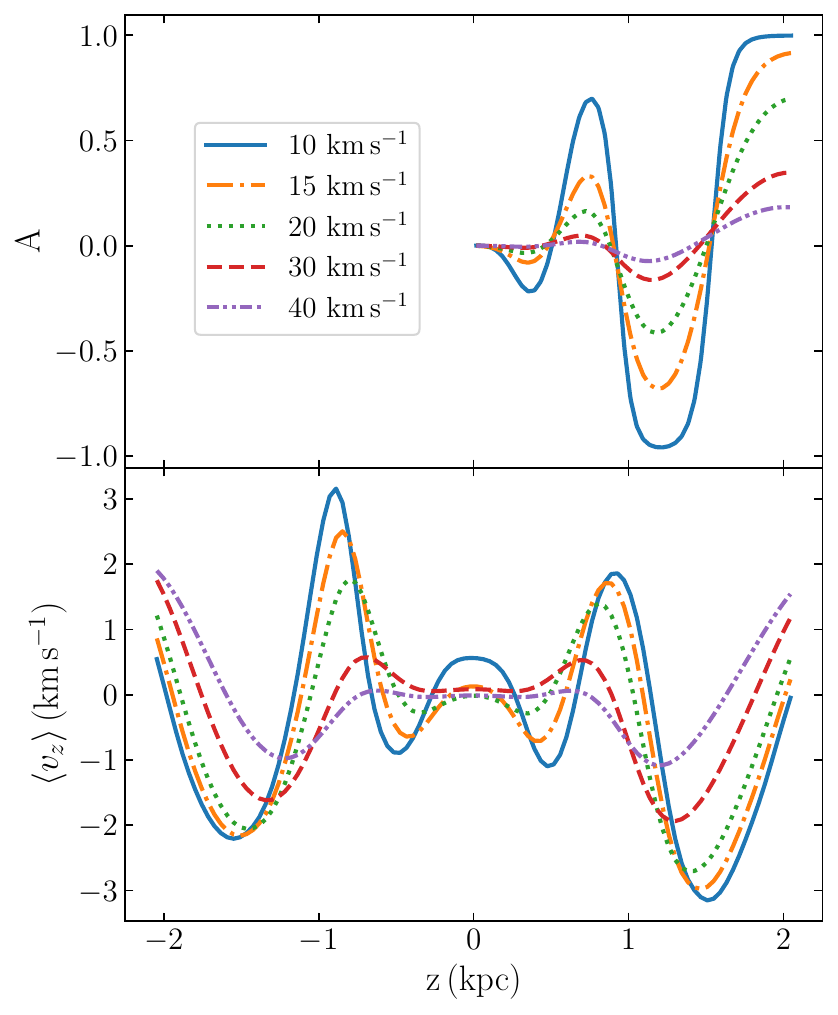}
    \caption{Effect of the disc's velocity dispersion on the perturbation from the example satellite perturbation. \textit{Top:} Asymmetry for five different isothermal discs set to have the same mid-plane density and varying velocity dispersions at $t=0$. \textit{Bottom:} Mean velocity as a function of height for the same set of models. A colder disc reacts more strongly to a perturbation than a warmer disc, which is especially noticeable in the asymmetry.}
    \label{fig:VD_pert}
\end{figure}

Beyond how the mass of the disc affects the asymmetry, we also want to investigate how the choice of tracers of the disc changes the observed asymmetry. This requires a different set-up than the investigation into the mid-plane density, because the tracers are all orbiting in the same potential but may have a different velocity dispersion. To reflect this, we integrate all of the orbits of the disc in the same isothermal potential with a mid-plane density of $\rho_o= 0.1\,\,\mathrm{M_\odot\, pc^{-3}}$ and velocity dispersion of $\sigma=20.5\,\, \mathrm{km\,s^{-1}}$. We then look at a set of five quasi-isothermal distribution functions which will represent the different tracers. The velocity dispersions are $\sigma=$ 10, 15, 20, 30, and 40 $\mathrm{km}\,\mathrm{s}^{-1}$. These velocities are chosen to reflect the span of velocity dispersion seen in the solar neighbourhood \citep[e.g.][]{ted}. 

In the top panel of \figurename \ref{fig:VD_pert}, we see the asymmetry for the five tracers with varying velocity dispersion. Clearly, as we increase the velocity dispersion, the magnitude of the asymmetry decreases. The fact that the change we see is mostly in amplitude and not the perturbation wavelength is not surprising. Each of the tracers are orbiting in the same overall potential, it is just their specific distribution function which is changing, but there is no change in the vertical frequency of the disc so we would expect the perturbation to have a similar perturbation wavelength for all tracers. Normalizing the asymmetry such that the last peak has the same magnitude for each distribution function (not shown) demonstrates that as we increase the velocity dispersion, the ratio of inner peak height to outer peak height decreases slightly. This means that as you look at higher velocity dispersion, not only is the overall amplitude of the asymmetry smaller, but the signal nearer the mid-plane is also weaker compared to the signal further out. There is also a small, almost imperceptible, shift in the wavelength  of the asymmetry, but it is nothing compared to the shift in the perturbation wavelength that we see in \secname \ref{sec:mpd}. We see similar effects in the vertical mean velocity in the bottom panel of \figurename \ref{fig:VD_pert}. The amplitude changes drastically, while the perturbation wavelength changes only minutely, although more noticeably than the asymmetry. 

Combining the results of this \secname and the previous one tells us that the wavelength  of the asymmetry is more dependent on the potential of the disc as opposed to what tracer we have chosen in the disc. We also found that the amplitude of the asymmetry is more sensitive to the velocity dispersion of the disc than it is to the mid-plane density. This could be an interesting tool to study the properties of the disc itself. Unfortunately, both the amplitude and the wavelength  of the asymmetry are also very dependent on the form of the perturbation, including how long ago it occurred, the mass of the satellite, and the velocity of the satellite as it passed through the mid-plane of the disc. This will be discussed further in \secname \ref{sec:sgr_orbit} and \secname \ref{sec:sgrA} when we examine the orbit of Sagittarius and the resultant asymmetries. 

\subsection{Number of Pericentres}\label{sec:peri}

One other area of investigation is how the number of times the satellite has passed through the disc affects the observed asymmetry. To test this, we use the satellite orbit and potential discussed in \secname \ref{sec:model_setup} and integrate it backwards through one, two and three pericentres so that it is still beginning and ending at apocentre.

\begin{figure}
    \centering
    \includegraphics[width=0.4\textwidth]{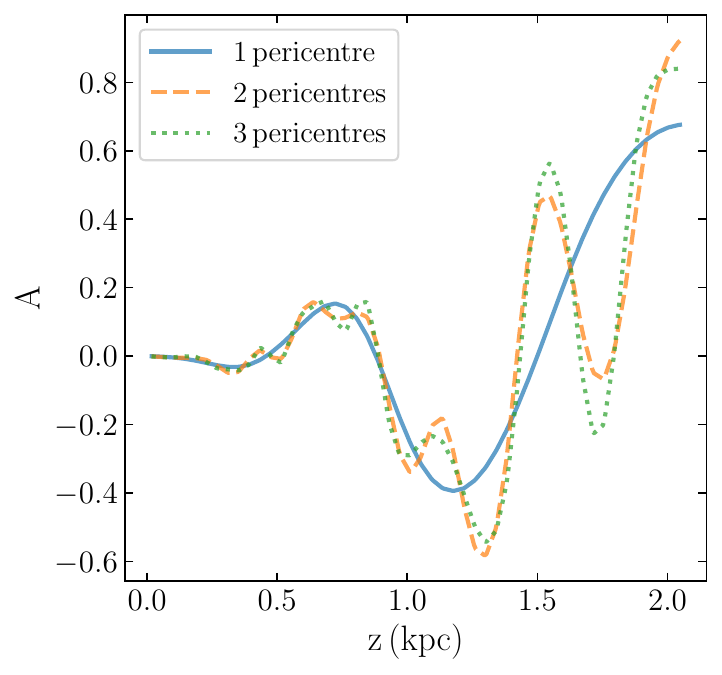}
    \caption{Model of the vertical number count asymmetry of the disc when we integrate the example Sgr orbit back through one (blue solid), two (orange dashed) and three (green dotted) pericentres. The three are similar out to 1 kpc at which point the one pericentre orbit case starts to diverge from the two and three pericentre cases. The two and three pericentre cases remain similar out to 2 kpc.}
    \label{fig:apo_A}
\end{figure}

A comparison of the three different resulting asymmetries is shown in \figurename \ref{fig:apo_A}. From this, we can clearly see that within 1 kpc, the asymmetry is similar for all three cases, so going back one pericentre is enough to capture the behaviour of the asymmetry. However, as we go further out, we see that one pericentre is not enough to capture the shape of the asymmetry. Further out than 1 kpc, the difference between the orbit with two and three pericentres is not significant. From this, we can conclude that to examine the asymmetry as far out as 2 kpc, an orbit with two pericentres is sufficient to see the overall shape of the asymmetry. While three pericentres does provide more detail, it also requires we go back further in time. As we go further back in time, the orbits become more uncertain and therefore our confidence in the resultant asymmetry also decreases. Therefore, when we calculate the asymmetry observed now due to a satellite in \secname \ref{sec:real}, we integrate our orbits back through two pericentres as a balance between accuracy and detail. We also do not include mass loss in our model of Sgr which will affect the amplitude of the additional perturbations from the second and third pericentre passages. In \citet{SgrModel} they found in their N-body simulations that Sgr lost a factor of $2-5$ of its mass after the penultimate pericentre and became gravitationally unbound after the most recent pericentre. This could mean that the difference in the asymmetry between one, two, and three pericentres passages would be even smaller in reality.

\section{One-Dimensional Simulations}\label{sec:1DSim}

To validate our model, we compare it to one-dimensional simulations using the \texttt{wendy} Python package, described in Appendix \ref{sec:wendy}. The benefit of a one-dimensional simulation is twofold. First, it allows for a direct comparison to our model. Second, it is computationally quick because one only needs the order and not distances for a one-dimensional simulation, as the gravitational force in one dimension is independent of distance. 

\subsection{Setup of the 1D simulations}\label{sec:1D_setup}

For our one-dimensional simulations, there are several things to consider. The first is how we initialize the equilibrium disc. Second, we consider the perturbation to the disc given by our satellite passage. Finally, we need to consider the amount of self-gravity in the disc. To achieve a large degree of accuracy in the density and velocity distribution resulting from the simulation, we use $N=10^7$ particles in our simulations.

To compare to our simple model from \secname \ref{sec:simple_model}, we want to initialize our disc as a single component quasi-isothermal disc with mid-plane density of $\rho_0=0.1\,\mathrm{M_\odot\,pc^{-3}}$ and a velocity of dispersion of $\sigma=20.5\,\mathrm{km\,s^{-1}}$. One method of doing this is to have particles which are equal in mass and the number of particles at each $z$ position match the quasi-isothermal disc distribution function. The problem with this is that as one moves further out from the mid-plane of the disc, the bins in $z$ contain less and less particles and Poisson noise begins to dominate the signal. Instead, we choose to draw $z$ positions from a uniform distribution between $-2$ and $2$ kpc and $v_z$ velocities from a uniform distribution between $-120$ to 120 $\mathrm{km\,s^{-1}}$. We choose these ranges to remain consistent with our model parameters. The uniform distribution ensures that when we look at the density at each timestep, each bin has roughly equal numbers. Next, we use these positions to calculate the action of each particle using the \texttt{IsothermalDiskPotential} from \texttt{galpy} like we did for our simple model. Using the action, we can assign a mass based on the value of the quasi-isothermal distribution in \eqnname (\ref{eq:qi_df}) at that point in phase-space. Without the different masses, the disc would be a uniform distribution of particles, but when weighted by their mass, the distribution function is quasi-isothermal. We then normalize the masses such that they add to one for units consistent with \texttt{wendy}. 

For the perturbation, we use the vertical force from a satellite as defined in \eqnname \eqref{eq:Fz}. We use both the orbit and the potential from \secname \ref{sec:simple_model}, where the orbit is from apocentre to apocentre and the Sgr potential has a total mass of $10.2\times10^9\,\mathrm{M_\odot}$ which includes both a stellar and dark matter component. We also include the rotation of the Sun around the Galactic centre in our calculation of the force like we did in \secname \ref{sec:simple_model}. These choices mean that the force in our simple model is the same as the force applied to the disc in our one-dimensional simulation. 

For our consideration of self-gravity, we use a method similar to \citet{widrow_alpha}. We consider an $\alpha$ parameter in our simulations which determines the strength of the self-gravity in the simulation. We do this by splitting the disc potential $\Psi$ into two parts, a self-gravitating part $\Psi_\mathrm{self}$ and a fixed part $\Psi_\mathrm{fixed}$, giving $\Psi = \alpha \Psi_\mathrm{self} + (1-\alpha)\Psi_\mathrm{fixed}$. In practice, we achieve this by multiplying the masses of the simulation particles by $\alpha$ and adding an external potential $(1-\alpha)\Psi_\mathrm{fixed}$ to the simulation. So, when $\alpha=1$, the simulation is fully self-gravitating and as $\alpha\rightarrow0$, the simulation approaches free particles in a fixed potential. In \secname \ref{sec:alpha}, we discuss the effects of changing the self-gravity in the disc.

At each time step we calculate the density, mean velocity, and their respective uncertainties. We choose 201 bins along the $z$-axis when calculating both the density and the velocity. The density is calculating by summing the masses of all the particles within each bin and dividing by the bin width, $\Delta z=19.9$ pc. We calculate the mean vertical velocity for the $j$th bin using:
\begin{equation}
    \langle v_{z}\rangle_j= \frac{\Sigma_i^{n_j} \,v_{z,i,j}\cdot m_{i,j}}{\Sigma_{i}^{n_j} \,m_{i,j}} \label{eq:vz}
\end{equation}
where $n_j$ is the number of particles in the $j$th bin and $v_{z,i,j}$ and $m_{i,j}$ are the vertical velocity and mass of the $i$th particle in the $j$th bin respectively. Practically, this means that if we multiply the numerator and denominator by the bin width, we only have to calculate the numerator at each step, because the denominator will become the density of the bin, which we already have. The uncertainties of both are calculated by summing the square of the weights in each bin.

Once we have recovered the density, we also calculate the mid-point of the density distribution by fitting a $\mathrm{sech}^2$ profile to the density and use that to recover the offset of the disc from zero. The effect of this fit is to remove an overall linear factor from the asymmetry and removes any `tilt', but does not affect the shape. We then recalculate the density adjusted for this offset by subtracting it from the positions of the particles. This is similar to the method used in \citet{ME} to adjust for the height of the Sun, except that ours is a single component disc, instead of a two component disc. With the density and mean vertical velocity of the disc as a function of time, as well as their uncertainties, we are equipped to test our model using the one-dimensional simulation.

\subsection{Basic tests of the linear-perturbation model}\label{sec:simvmodel}

\begin{figure}
    \centering
    \includegraphics[width=0.5\textwidth]{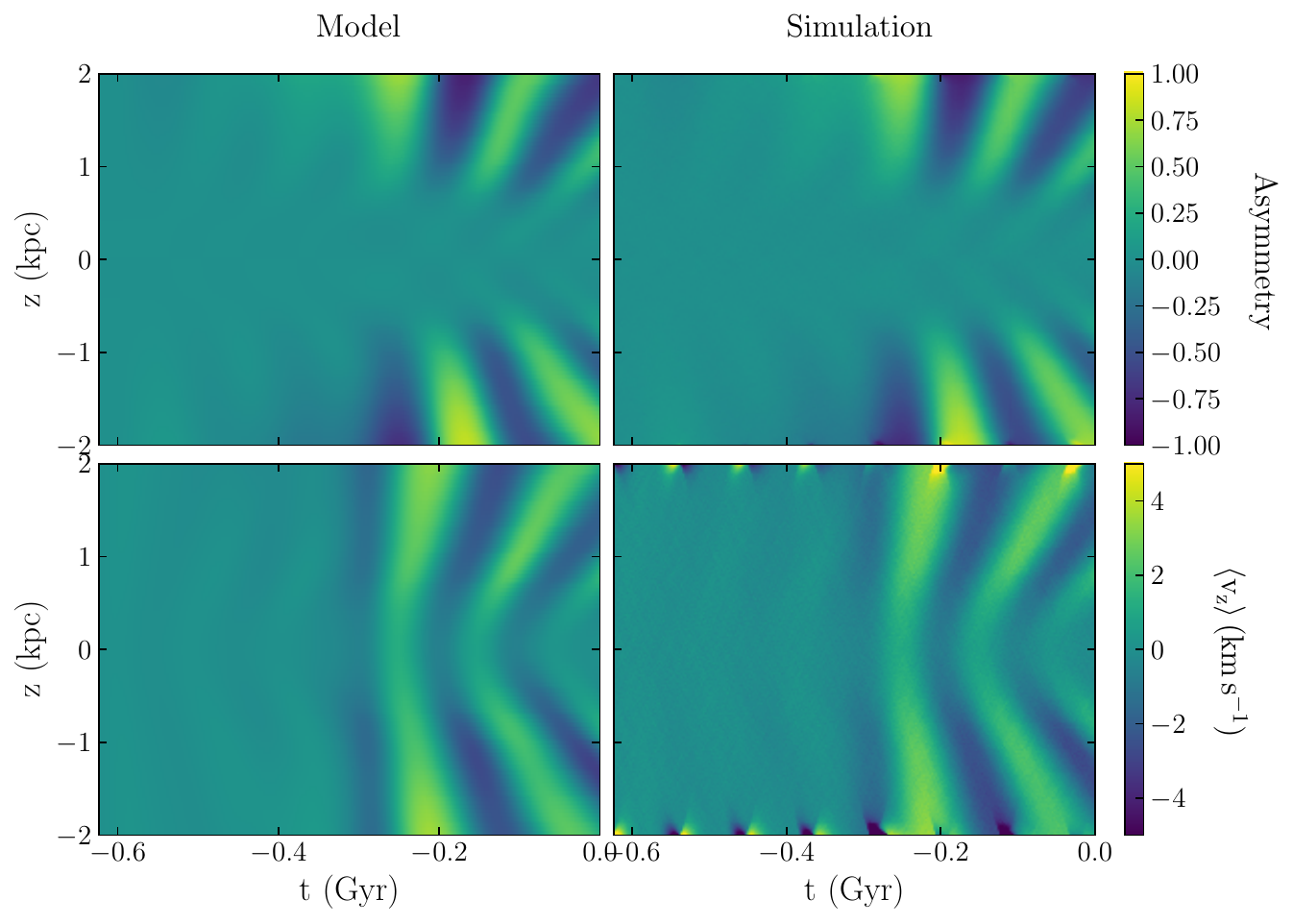}
    \caption{Comparing the linear-perturbation model to one-dimensional simulations. In our simulation, $\alpha$ parameterizes the self-gravity in the disc where $\alpha=0$ has no self-gravity and $\alpha=1$ is fully self-gravitating. \textit{Top:} Asymmetry as a function of time and height above the mid-plane for the model (left) and the one-dimensional simulation with $\alpha=0.01$ (right). \textit{Bottom:} the mean velocity as a function of time and height above the mid-plane for the model (left) and the simulation with $\alpha=0.01$ (right). The perturbation to both is from the orbit and satellite potential discussed in \secname \ref{sec:simple_model}. The simulation is well matched by the model, both in amplitude and the perturbation wavelength.}
    \label{fig:MvsS}
\end{figure}

As validation of our model, we compare it to the one-dimensional simulation. Our model does not account for self-gravity, so it should yield nearly the same results as the case where our self-gravity parameter, $\alpha$ approaches zero. We do not expect the model to match the simulation exactly, even in the $\alpha=0$ limit, because \texttt{wendy} calculates the non-linear perturbation along the perturbed orbit whereas the model is calculating the linear perturbation approximation along the unperturbed orbit. So long as the perturbation is sufficiently small, we expect the difference between the two calculations to be minor and should see good agreement between the model and the simulation. Since \texttt{wendy} cannot handle massless particles, we instead consider the case where $\alpha=0.01$ which should be small enough to be negligible and is considered a minimally self-gravitating scenario. It should be roughly equivalent to the $\alpha=0$ case and therefore approximately equivalent to our model if our assumptions are correct.

In both the model and the simulation, we have fit the density at each time step to recover the shift of the density from the mid-plane. This is akin to fitting for the height of the Sun, $z_\odot$, when dealing with \emph{Gaia} data. This technique will allow us to compare the model to the asymmetry detected in \emph{Gaia} DR2 in \secname \ref{sec:real} where we look at more realistic orbits. After fitting for the mid-plane location, we find that the asymmetry for $|z|<0.5$ kpc becomes very small, and we only really see a signal outside of that range in heights.

\figurename \ref{fig:MvsS} shows a comparison between our model and the 1-dimensional simulation where the disc is minimally self-gravitating. Clearly the simulation and model are an excellent match. The top row shows the asymmetry as a function of time and $z$. the shape, position, and amplitude of the oscillations agree between the model and simulation. Even the ridges before the pericentre passage of the satellite matches between the two. The bottom row shows the mean vertical velocity as a function of $z$ and time.  For the most part, the mean velocity in the model and the simulation agree well. When $\left|z\right|>1.85$ kpc, the velocity in the simulation shows a signal which clearly does not match the model. This is because of the method we use to calculate the mean velocity in the simulation as discussed in \eqnname \eqref{eq:vz}. As we move away from the mid-plane of the disc, the density is much smaller and the error begins to dominate more. This means that further out in the simulation, the error is dominating our mean velocity measurement. Other than this, the model and one-dimensional simulation are in great agreement for the mean vertical velocity as well as the asymmetry.

Finally, to follow up the investigation of \secname \ref{sec:peri}, we looked at the difference between one, two, and three pericentre passages in the simulation. With $\alpha=0.01$, the simulation behaves exactly as the model behaved. This is unsurprising, but works to confirm that our model is successfully reproducing the behaviour of a minimally self-gravitating disc. The next step is to see how well it reproduces the asymmetry once we include self-gravity in our simulation. 

\subsection{Importance of self-gravity}\label{sec:alpha}

\begin{figure*}
    \centering
    \includegraphics[width=0.95\textwidth]{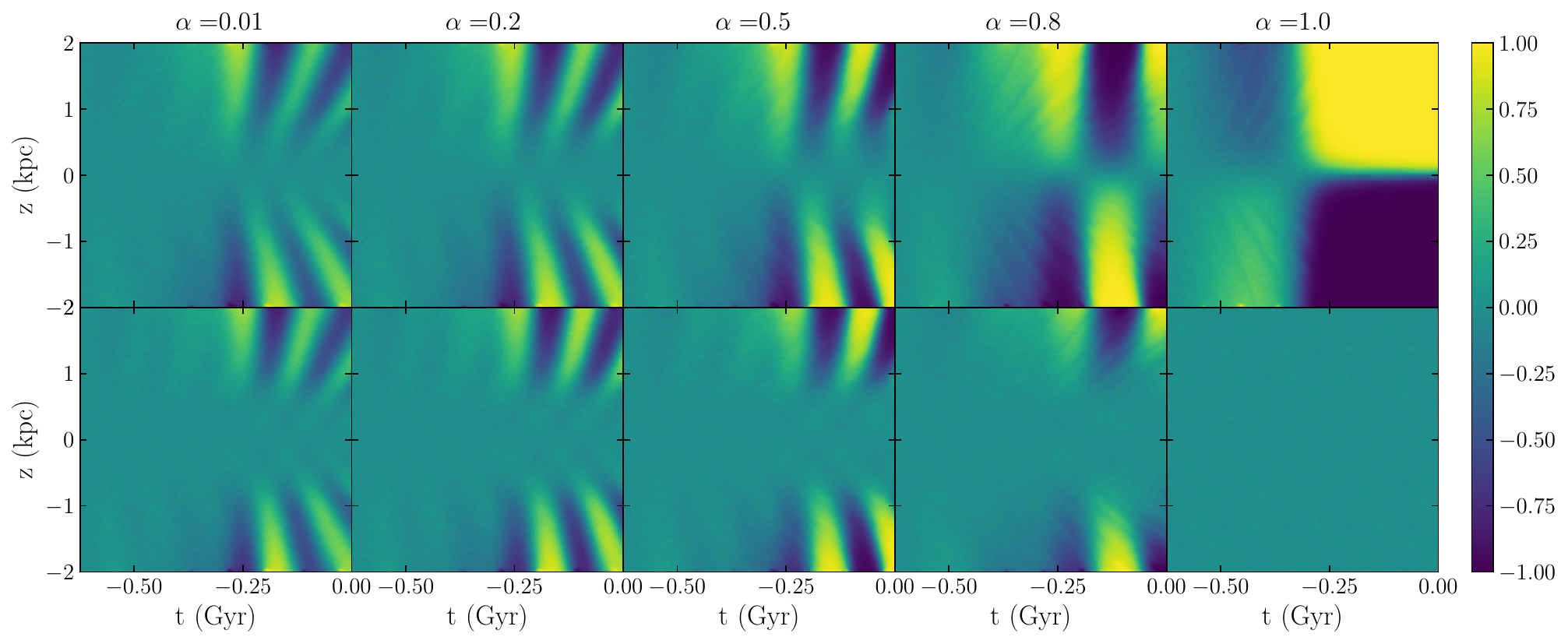}
    \caption{Effect of self-gravity on the asymmetry resulting from a satellite perturbation. \textit{Top:} Asymmetry as a function of height above the mid-plane and time for the one-dimensional simulations. From left to right, we increase the self-gravity in the simulation, with $\alpha=1$ corresponding to a fully self-gravitating disc. 
    \textit{Bottom:} Same as the top panel, but corrected for the perturbed location of the mid-plane of the disc. As we increase the self-gravity, both the amplitude and the wavelength  of the asymmetry increases. At $\alpha=0.8$ and higher, the stiffness of the self-gravitating disc causes the simulation to stop resembling the model.}
    \label{fig:alpha}
\end{figure*}

In this subsection, we investigate the effect of self-gravity on the observed asymmetry using 1-dimensional simulations. To do this we consider the five cases of $\alpha=0.01,0.2,0.5,0.8$ and $1.0$. We choose 0.01 and 1.0 because these represent the case of a minimally self-gravitating disc and fully self-gravitating disc respectively. The other three $\alpha$ values were chosen because they were used by \citet{widrow_alpha} in their investigation of eigenfunctions of the Galactic phase-space spirals. In their work, they found that $\alpha=0.8$ did not allow for enough phase-mixing to result in spiral structure in phase-space and that the moderate amounts of self-gravity, $\alpha=0.2$ or $0.5$ resulted in spirals, and were therefore likely to be a better representation of the solar neighbourhood. One might expect that $\alpha$ for the solar neighbourhood disc might be 1, since the disc is self-gravitating. However, this is only true in the 3-dimensional case. In the 3-dimensional case, there are unperturbed stars at different radii and azimuths which impose a force on the stars at the solar neighbourhood that can be represented by a fixed potential. The potential of the dark matter in the solar neighbourhood will also contribute to the fixed portion of the potential. These outside factors do not exist in a 1-dimensional simulation, which is why we consider different self-gravity scenarios in our 1-dimensional simulations.

\figurename \ref{fig:alpha} shows the asymmetry as a function of time for the five different values of $\alpha$. The top row shows the asymmetry before we have fit for the mid-plane of the density and the bottom row shows the asymmetry after adjusting for that. Similar to what was observed in \secname \ref{sec:simvmodel}, we find that for all values of $\alpha$, the asymmetry within $|z|<0.5$ kpc becomes very small after fitting for the location of the mid-plane. As we increase the self-gravity of the disc, we see an increase in the amplitude of the asymmetry and a decrease in the perturbation wavelength. We also see that before fitting for the location of the mid-plane, the asymmetry becomes less of an oscillation and begins to resemble a dipole, reflecting the behaviour of a much more rigid disc. This is because as the satellite passing by exerts a force on the disc, the simulations with a weaker fixed component of the potential are dragged along by the passing of the satellite. However, in the bottom panel once we have fit for the mid-plane of the density, we see this signal disappear and the asymmetry is again a clear oscillation. We even see that the case with full-self gravity, $\alpha=1$, actually has the weakest asymmetry signal; a fully self-gravitating disc is highly rigid and does not start oscillating due to a perturbation.

Obviously, the asymmetry of the model most closely resembles the minimally self-gravitating case. However, if we look at the asymmetry at the final time, there are some notable trends. When $\alpha=0.2$, the overall shape of the asymmetry remains similar, but the amplitude and perturbation wavelength both increase. We also see that the location of the peaks shift inwards towards $z=0$. For the $\alpha=0.5$ case, \figurename \ref{fig:alpha} shows that both the perturbation wavelength and the amplitude of the ridges have increased compared to the two smaller $\alpha$ cases. Additionally, the angle of the ridges decreases. So, while the asymmetry looks similar overall, the combination of these three effects makes the asymmetry at a given time look quite different than the model. While the sign of the asymmetry typical matches the model, the decrease in the angle of the ridges means that there are less oscillations in the asymmetry at a given time and the increased amplitude is significant compared to the increase between the $\alpha=0.01$ and $\alpha=0.2$ case. When $\alpha=0.8$, the simulation has lost all similarity to the model. In \figurename \ref{fig:alpha}, we see that for $\alpha=0.8$, there are less ridges than seen in the model. This means that for a large portion of the time, the asymmetry in the simulation has a different sign than the model. Similar to the $\alpha=0.5$ case, we also see a decrease in the angle of the ridges, making the asymmetry at a given time appear monotonic. Finally, when we consider $\alpha=1.0$, fitting for the shift of the mid-plane entirely removes the signal of the satellite. In this case, the disc is simply dragged around by the satellite, but does not experience much of an internal perturbation. 
\begin{table}
\centering
\begin{tabular}{c|cccc}\hline
              & \multicolumn{2}{c}{$\mathbf{|z|<0.5}$ \bf{kpc}} & \multicolumn{2}{c}{$\mathbf{0.5\,\mathrm{\bf{kpc}}<|z|<1.9}$ \bf{kpc}} \\
              &$\mathbf{\langle\left|\Delta A\right|\rangle}$&    $\boldsymbol{\%}_\mathbf{{\left|\Delta A\right|>0.07}}$&  $\mathbf{\langle\left|\Delta A\right|\rangle}$&    $\boldsymbol{\%}_\mathbf{{\left|\Delta A\right|>0.07}}$ \\ \hline\hline
$\alpha=0.01$ & 0.012&  0.004 & 0.019&  2.0         \\
$\alpha=0.2$  & 0.013&  0.034 & 0.029&  19.3          \\
$\alpha=0.5$  & 0.015&  0.335 & 0.046&  27.9          \\
$\alpha=0.8$  & 0.014&  0.080 & 0.046&  28.0          \\
$\alpha=1.0$  & 0.014&  0.078 & 0.056&  31.9         
\end{tabular}\label{tb:alphaerr}
\caption{Median difference between the asymmetry in the model and the simulation. Percentage of bins where the difference between the asymmetry in the model and the simulation is greater than 0.07. Both for different amounts of self-gravity,$\alpha$, and ranges of $z$}
\end{table}

To compare the five simulations with different values of $\alpha$, we consider two different statistics. For each of these statistics, we also consider two ranges of distance from the mid-plane. The first range is $|z|<0.5$ kpc, because we have already seen that the asymmetry is small in this range in all cases. The second bin is $0.5\,\mathrm{kpc}<|z|<1.9$ kpc. We chose an upper bound of 1.9 because similar to the velocity discussed in \secname \ref{sec:simvmodel}, the uncertainty in the asymmetry on the extremities of our $z$ range is larger. The first statistic we consider is the median absolute difference between the model asymmetry and the asymmetry in the simulation for each $\alpha$ value, $\langle\left|\Delta A\right|\rangle$. The second statistic is the percentage of bins where the magnitude of the difference is above a threshold of 0.07, $\%_{\left|\Delta A\right|>0.07}$. We chose this value because it corresponds to approximately 10\% of the maximum asymmetry. The calculated values for each $\alpha$ simulation and each distance range are shown in \tablename \ref{tb:alphaerr}. We see that the median difference and the percentage of bins above 10\% across all simulations is fairly consistent for the range $|z|<0.5$ kpc. Both measures of the difference between the simulation and the model are also consistently small across all values of $\alpha$. This is because the asymmetry is small across all simulations for that range of heights. For the range $0.5\,\mathrm{kpc}<|z|<1.9$ kpc, both the median difference and the percentage increase drastically as we increase $\alpha$. These values confirm our previous assertion that $\alpha=0.01$ is a good fit to the model, $\alpha=0.2$ is a somewhat match to the model, and $\alpha=0.5,\,0.8,$ and 1.0 are very different from our model.

Finally, we looked at comparing multi-pericentre passages of the satellite again, but this time with $\alpha=0.2$. The difference in the asymmetry between an orbit which goes through one, two, and three pericentres actually turns out to be less pronounced than in the model or the $\alpha=0.01$ case. This is likely because as you introduce self-gravity, the oscillations due to the perturbation are more quickly damped out. Meaning that the earlier pericentre passages are less important in the self-gravitating case. This confirms our decision in \secname \ref{sec:peri} to only go back through 2 pericentres.

\section{Comparing to Sagittarius}\label{sec:real}

In this section, we use the modeling machinery developed above to answer the question of whether the perturbation to the local solar neighbourhood due to the passage of the Sgr dwarf galaxy can be the cause of the observed asymmetry in the vertical number counts, the bending-like perturbation to the mean vertical velocity, and the \emph{Gaia} phase-space spiral.

\subsection{Mass models of the Sgr dwarf galaxy}\label{sec:sgr_model}

As part of our model, we need to calculate the force due to a perturber on the solar neighbourhood to calculate the perturbation to the disc. If we assume that the perturber is the Sagittarius dwarf galaxy, then we need to consider two main things: what mass model describes the Sgr dwarf and what is its orbit. To calculate the orbit with dynamical friction, we need to first decide on the mass model of the satellite. In \citet{SgrModel}, they study the 3D structure and kinematics of Sgr. One aspect of their investigation includes running $N$-body simulations of the evolution of Sgr over the past $2-2.5$ Gyr. For their simulations, they considered several different initial conditions for the size and shape of Sgr. Each model has two components, stellar and dark matter, where the stellar component makes up approximately 10\% or less of the total mass within a few kpc. For most models, they have a rotation curve which peaks at 3 kpc with a velocity of approximately 50 to 60 $\mathrm{km\,s^{-1}}$. This corresponds to a total mass of approximately $(1$ to $3)\times 10^9\,M_\odot$ and a scale radius of $a_\mathrm{Sgr}\approx 3$ kpc. They also find that the progenitor today has a total mass of approximately $4\times10^8\,\mathrm{M_\odot}$, where the stars make up roughly a quarter of that mass and the rest is dark matter.

We use the above parameters when choosing our models for our investigation of both the Sgr orbit and the resultant perturbation to the disc. We consider a wide range of mass models that covers the heaviest initial conditions, down to a total mass less than the current day progenitor. As is commonly done, we choose to represent Sgr by two Hernquist profiles \citep{Laporte2018,SgrModel}: one for the stellar component, characterized by the mass, $M_*$ and scale radius, $a_*$, and one for the dark matter component characterised by a total mass, $M_\mathrm{DM}$, and scale radius, $a_\mathrm{DM}$. The values of these parameters for each of the models are shown in \tablename \ref{tb:Sgr_model}. We chose these structural parameters by matching the model specifications described above from \citet{SgrModel} for Sgr 2. We then consider one model which is heavier and three lighter models to cover the full range of possible Sgr masses. The mass of the stellar component for Sgr 2 is calculated to match their models as well. From there, we scale the stellar masses by the ratio of the DM masses, and we scale the radii of both the stellar component and the dark matter component by the square-root of the ratio of the masses.

\begin{table}
\centering
\begin{tabular}{|c|c|c|c|c|}\hline
    \textbf{Model} & $\bf{M_*\,}$ & $\bf{a_*}$ & $\bf{M_{DM}}$ & $\bf{a_{DM}}$\\ 
     & $(10^9\,\mathrm{M_\odot)}$ &(kpc)&$(10^9\,\mathrm{M_\odot})$&(kpc)\\\hline\hline
     Sgr 1 & 1 & 1.5 & 50 & 6.7  \\
     Sgr 2 & 0.2 & 0.65 & 10 & 3.0 \\
     Sgr 3 & 0.1 & 0.46 & 5 & 2.1 \\
     Sgr 4 & 0.02 & 0.21 & 1 &  0.95 \\
     Sgr 5 & 0.01 & 0.15 & 0.5 & 0.67
\end{tabular}
\caption{Properties of the stellar and dark matter components of the five different Sagittarius models.}
\label{tb:Sgr_model}
\end{table}

\subsection{The Orbit of Sgr}\label{sec:sgr_orbit}

\begin{figure}
    \hspace{-0.6cm}
    \includegraphics[width=0.52\textwidth]{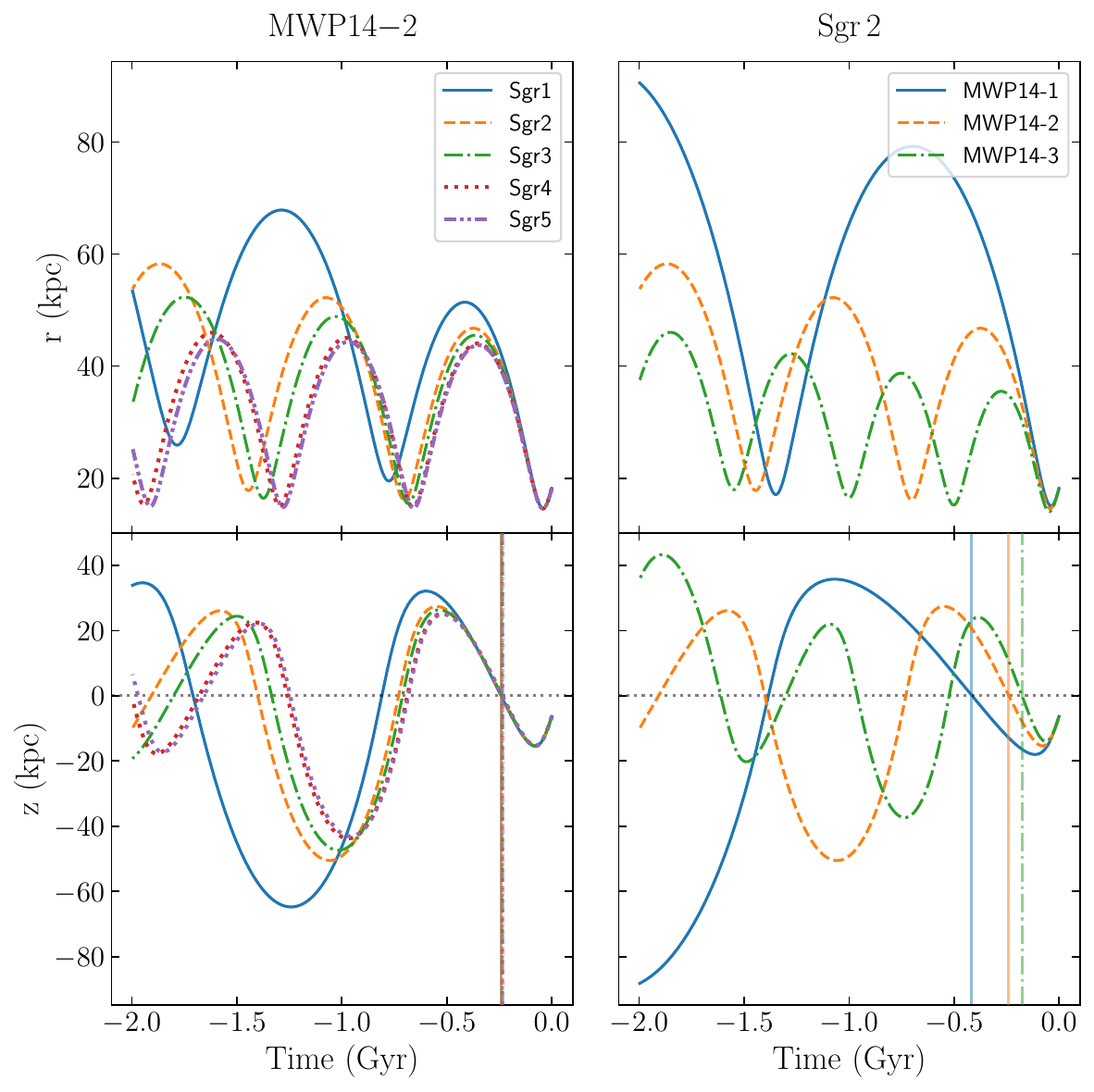}
    \caption{The variety of Sgr models consistent with its observed present-day position. \textit{Left: }An example of a Sgr orbit over the past 2 Gyr for four different models of Sgr. Their properties are shown in \tablename \ref{tb:Sgr_model}, but their total masses in order are $(51,10.2,5.1,1.02,0.51)\times10^9\mathrm{M_\odot}$. Shown are the Galactocentric radius with time (top) and the vertical position of Sgr as a function of time (bottom). The vertical lines in the bottom plot are $t_{\mathrm{through}}$, the point at which Sgr most recently passed through the plane and when $v_{z,\mathrm{through}}$ is calculated. They all occur at very similar times for the four different Sgr models. 
    \textit{Right: }An example of a Sgr orbit over the past 2 Gyr in the three different Milky Way potentials. Where MWP14-1 (blue solid line) is a Milky Way potential with a `light' halo, MWP14-2 (orange dashed line) is the `medium' halo, and MWP14-3 (green dash-dot line) is the potential with a `heavy' halo. Shown are the Galactocentric radius with time (top) and the vertical position of Sgr as a function of time (bottom). Again, the vertical lines are $t_{\mathrm{through}}$, the point at which Sgr most recently passed through the plane and when $v_{z,\mathrm{through}}$ is calculated.}
    \label{fig:Sgr_sample}
\end{figure}

\begin{figure*}
    \centering
    \includegraphics[width=\textwidth]{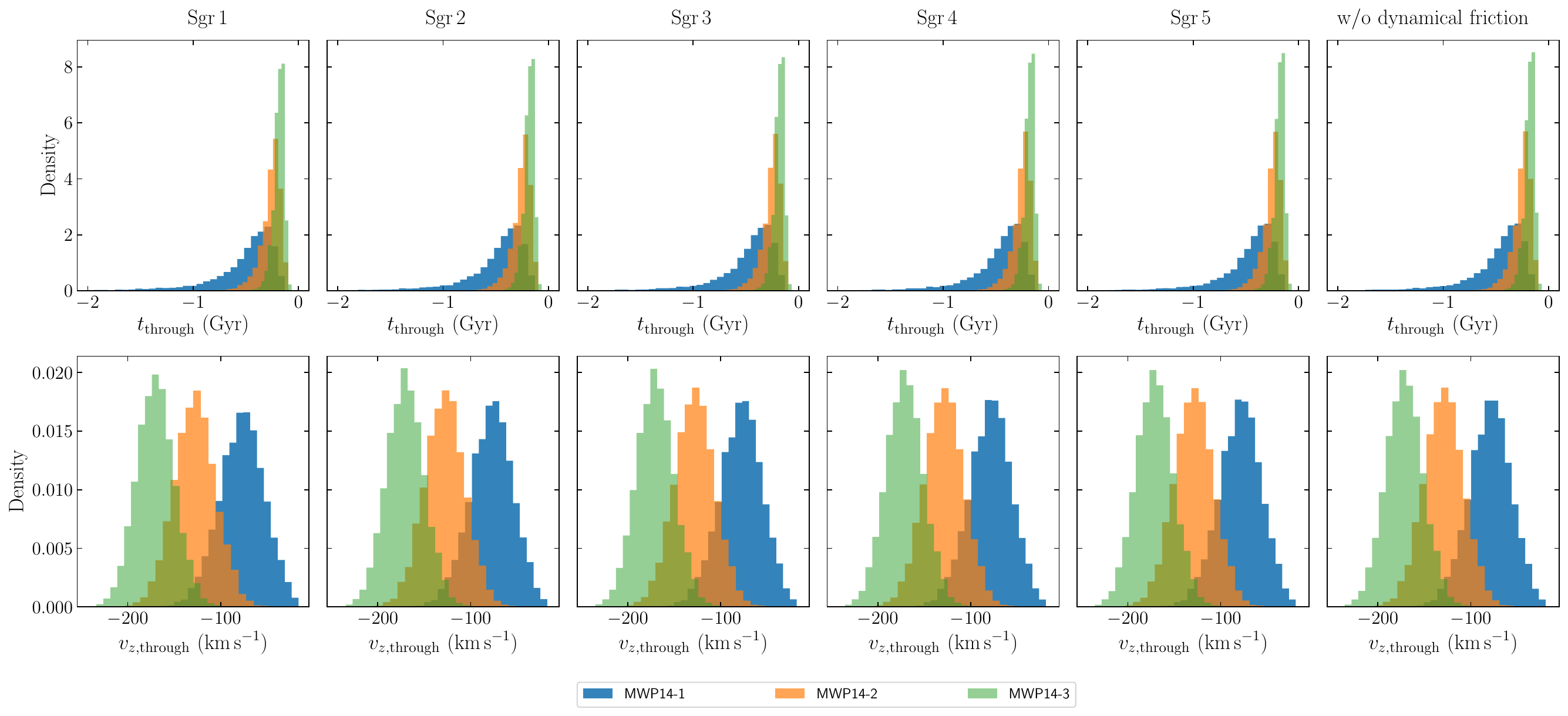}
    \caption{Impact properties of Sgr's last passage through the disc. \textit{Top:} Distribution of the time at which Sagittarius passed through the mid-plane of the disc, $t_{\mathrm{through}}$, for 5 different Sagittarius models and 3 different Milky Way models. The right most panel shows the case with no dynamical friction.
    \textit{Bottom:} Distribution of the velocity at which Sagittarius passes through the mid-plane, $v_{z,\mathrm{through}}$, for the same Milky Way and Sgr models.
    As we increase the mass of the halo, Sgr passes through the disc more recently and at a faster velocity.}
    \label{fig:through}
\end{figure*}

One of the important ingredients of our method is the orbit of our satellite which is needed to calculate the vertical force on the disc. If we want to be able to compare observations of the number count asymmetry and mean vertical velocity to a realistic model of the perturbation due to Sagittarius, we need to have a good idea of the orbit of Sgr. However, there is still some uncertainty surrounding the current position of the Sgr dwarf galaxy progenitor in 6-dimensional phase-space. Uncertainty in the exact current day position and velocity of Sgr lends a lot of uncertainty to the orbit of Sgr. Therefore, we chose to simulate several different orbits of Sagittarius to figure out the most likely parameters to use for our model and to explore the full range of perturbations that Sgr may have caused, depending on its exact orbit. We have already discussed different mass models of Sgr in \secname \ref{sec:sgr_model}, the uncertainty in these as well as the uncertainty in the Milky Way potential adds even more uncertainty to the orbit which is why we consider not only a range of initial conditions, but also different models for Sgr and the Milky Way.

To initialize the orbit, we use the positions, velocities, and corresponding uncertainties from \citet{GaiaSatKin} for the 5-dimensional parameter space of Sgr. For the distance to Sgr, we use $R= 26\pm 2$ kpc \citep{SgrDist}. We then sample a normal distribution centered at the values given in Table C.2 of \citet{GaiaSatKin} with a standard deviation given by the quoted errors in the same table. For robustness, we also consider different models for the Milky Way potential as well as the Sgr mass models discussed in \secname \ref{sec:sgr_model}. For the Milky Way potential, we consider three different potentials. The first is a `light' halo, MWP14-1, which is the Milky Way potential \texttt{MWPotential2014} from  \texttt{galpy} and has a halo mass of $8\times10^{11}\,\mathrm{M_\odot}$ \citep{galpy}. The second `medium' halo, MWP14-2, has the same disc and bulge potential as the first one, but the halo has been made 1.5 times heavier. Finally MWP14-3, the `heavy' halo is also the \texttt{MWPotential2014}, but with a halo which is twice as heavy. 

We calculate the orbits both with and without dynamical friction to compare the two scenarios. For both the simulations with dynamical friction and without, we sample the initial conditions 10,010 times from the Sgr phase-space and then integrated the orbits using \texttt{galpy} in each of the three different Milky Way potentials. In the case of no dynamical friction, we did not investigate the different Sgr potentials since the shape and weight of Sgr has no bearing on an orbit without dynamical friction. For the cases with dynamical friction, we integrate the initial conditions in the three different Milky Way potentials for each of the Sgr models used to calculate the dynamical friction component. \figurename \ref{fig:Sgr_sample} shows how the different Sgr mass models and Milky Way potentials effects the same initial conditions. All orbits in the figure start with the same position and velocity. On the left, we consider the case where the Milky Way potential is constant and we vary the Sgr mass model. From this, we see that the different Sgr mass models do not have a large effect on the time since Sgr passed through $z=0$. On the right, we use the Sgr 2 mass model and integrate it in different Milky Way potentials. Changing the mass of the Milky Way halo clearly has a much larger effect on the time since passing through the disc. 

We need decide on a measure to quantify the difference between the many orbits we consider. To compare the different orbits we chose two quantities that will reflect the impact of Sgr on the vertical component of the disc: the speed at which it passed through the plane defined by $z=0$, $v_{z,\mathrm{through}}$ and the time when it passed through the same plane, $t_{\mathrm{through}}$. \figurename\ref{fig:through} shows the distribution of $v_{z,\mathrm{through}}$ for each of the models described above with dynamical friction. 

The top row of \figurename \ref{fig:through} shows the trend in the time since passing through the mid-plane for the different Milky Way and Sagittarius models. As expected, by increasing the mass of the halo in the Milky way model, the time since passing through the disc decreases due to the tighter orbit explained above. The bottom row of \figurename \ref{fig:through} shows the distribution of the velocity of Sagittarius as it passes through the mid-plane most recently. The first obvious trend is that the speed in the case where we include dynamical friction is not significantly different than the cases where it is included. There is also no obvious trend in $v_{z,\mathrm{through}}$ as we look at the different Sgr models in each Milky Way model. The second obvious trend is that Sgr passes through the mid-plane of the disc at a much faster speed as we increase the mass of the Milky Way halo. This is unsurprising since increasing the halo mass also increases the mass inside Sagittarius' orbit which will then place it on a tighter orbit.We can now use these orbits to investigate the asymmetry and mean velocity trends due to Sgr on the solar neighbourhood. 

When we look at the 2D histogram of $t_{\mathrm{through}}$ and $v_{z,\mathrm{through}}$ (not shown), we find that the time since passing through $z=0$ and the velocity at which Sgr passed through are highly correlated. This is not surprising because the passage of Sgr through the mid-plane occurred relatively recently. This means that the spread in the initial condition parameters have less of an impact because there just isn't enough time for the orbits to diverge before passing through the disc. Now that we have the orbits for Sgr, we can look at the perturbation it causes in the disc.

\subsection{Investigating realistic disc and MW models}\label{sec:MWmodels}

\begin{figure}
    \centering
    \includegraphics[width=0.45\textwidth]{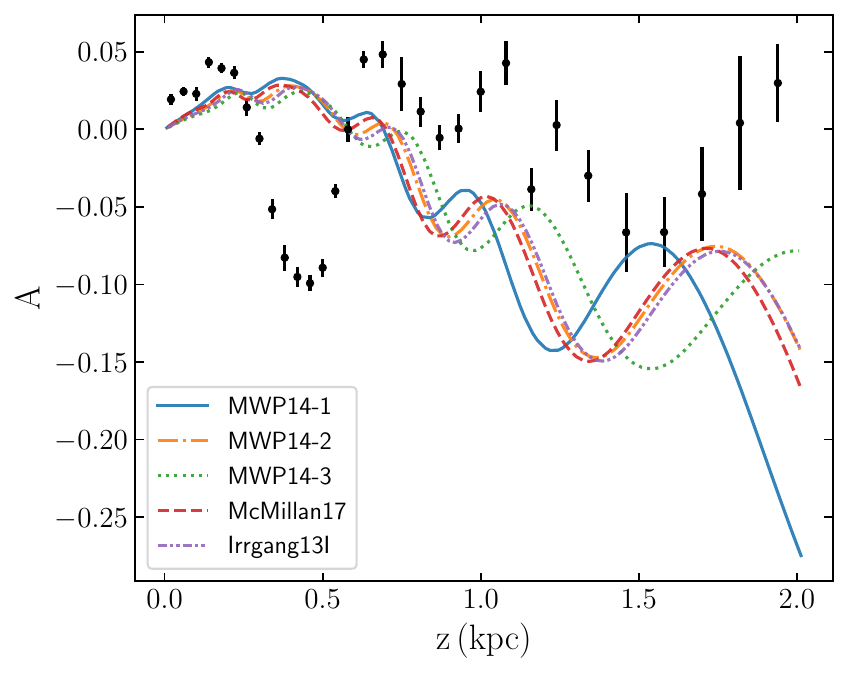}
    \caption{Effect of the disc potential on the density asymmetry caused by Sgr. The force from a Sgr orbit as a function of $z$ and $t$ is kept consistent for 5 different realistic Milky Way potential models. The orbit for Sgr goes back through two pericentres, starting at an apocentre and ending at the position of Sgr today and the Sgr mass model used is Sgr 2. The trend in the asymmetry is similar to that seen in \secname \ref{sec:mpd}. However, the model \texttt{McMillan17} does not behave as expected given its mid-plane density, because its more complex vertical structure also has a significant effect on the asymmetry. Within $|z| < 1$ kpc, the different disc models predict very similar asymmetries.}
    \label{fig:realMWComp}
\end{figure}

\begin{figure*}
    \centering
    \includegraphics[width=\textwidth]{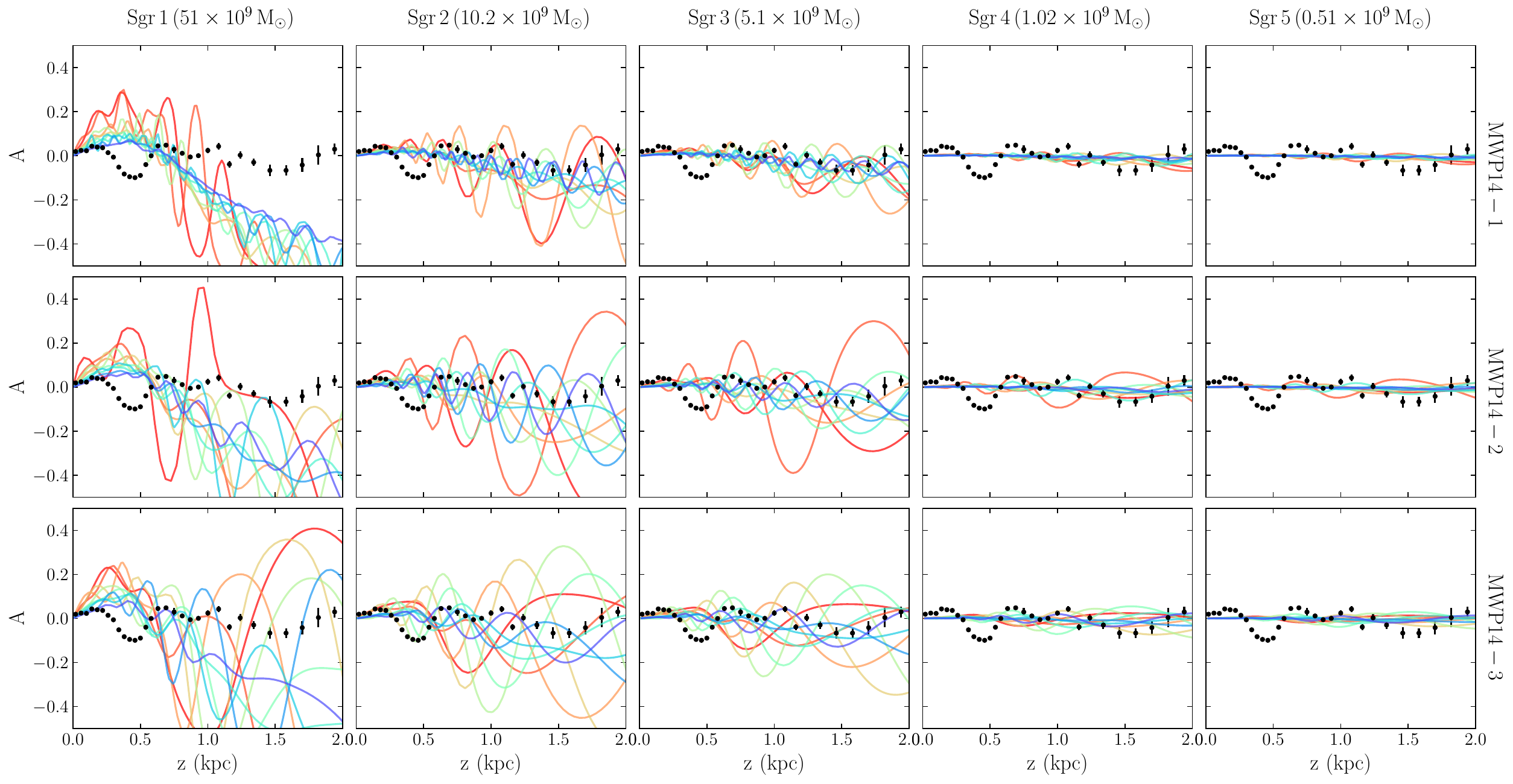}
    \caption{Sgr versus the data. Asymmetry for five different Sgr models and three different Milky Way models compared to the data from \citet{ME}. The masses of the different Sgr masses are printed at the top of the plot. MWP14-1 is the `light' Milky Way model, MWP14-2 is the `medium' one, and MWP14-3 is the `heavy' Milky Way model. Each plot has the model calculated for 10 representative orbits from the analysis of Sgr's orbit in \secname \ref{sec:sgr_orbit} from fastest through the mid-plane (red) to slowest through the mid-plane (blue). The black points in each plot are the observations from \emph{Gaia} DR2 taken from \citet{ME}.}
    \label{fig:Sgr_Asym}
\end{figure*}

In \secname \ref{sec:theory}, we considered a simple isothermal potential and distribution function. We then used that setup to see how changing the potential and the distribution function affects the results of the perturbation by a satellite. In this and the next sections, we want to look at a more realistic setup for the Milky Way. Specifically, in this section, we look at how more realistic Milky Way discs respond to the same perturbation before considering the full combinations of Milky-Way and Sgr models in the next section and comparing it to simulations. 

To do this, we take the orbit with the same initial conditions as \secname \ref{sec:simple_model}, but instead of integrating it from apocentre to apocentre, we integrate the orbit from two apocentres ago to the present-day position of Sgr using Sgr 2 as the Sgr mass model and MWP14-2 as the Milky Way potential in which the satellite is orbiting (shown by the orange dashed line in \figurename \ref{fig:Sgr_sample}). To do this, we define a force function which depends on vertical position and time only which can be used to calculate the vertical force on the disc for our model. To do this, we fix not only the satellite's orbit, but also the orbit of the solar neighbourhood around the centre of the Milky Way. We do this by choosing a constant circular velocity of $\mathrm{220\, km\,s^{-1}}$ for all five models. As the solar neighbourhood rotates around the galaxy, different circular velocities would affect the distance between the satellite and the solar neighbourhood as the satellite passes through the disc and would result in very different forces on the solar neighbourhood. The purpose of this section is to investigate the reactions of the five different realistic and complex discs to the same perturbation. In \secname \ref{sec:sgrA}, when comparing the effects of Sgr to the \emph{Gaia} DR2 data, we take the circular velocity of each Milky Way model into account. By applying this same force to different disc models, we are able to understand how more complicated potentials react to the same perturbation.

For the potential of the disc, corresponding to the unperturbed Hamiltonian, we look at five different realistic Milky Way potentials. The first three realistic potentials are the same three discussed in \secname \ref{sec:sgr_orbit}. The first potential, MWP14-1, is \texttt{MWPotential2014} from \texttt{galpy}, MWP14-2 is the same as MWP14-1, but with a halo which is 1.5 times heavier. The third potential, MWP14-3, is also \texttt{MWPotential2014}, but with a halo that is 2 times more massive. The final two potentials are \texttt{McMillan17} \citep{McMillan17} and \texttt{Irrgang13I} (model I from \citealt{Irrgang13I}) also taken from \texttt{galpy}. The \texttt{McMillan17} potential consists of a axisymmeterized version of a \citet{McmillanBulge} bulge, an exponential thin and thick stellar disc, an exponential \textsc{Hi} and molecular gas disc with holes in the centre, and a NFW dark matter halo \citep{NFW}. It also places the sun at $R_\odot=8.2$ kpc with a circular velocity of $232.8\,\mathrm{km\,s^{-1}}$. \texttt{Irrgang13I} includes a \citet{MN_model} bulge and axisymmetric disc, and a modified \citet{IrrgangHalo} spherical dark matter halo. It also places the Sun at $R_\odot=8.33$ kpc with a peak circular velocity of $242\,\mathrm{km\,s^{-1}}$. In \secname \ref{sec:mpd} above, we explored the effect of the mid-plane density of the disc, or equivalently, the vertical frequencies on the shape of the asymmetry. For our five potentials, the mid-plane densities and vertical frequencies are given in \tablename \ref{tb:pot_p0_vf}.

\begin{table}
\centering
\begin{tabular}{|c|c|c|}\hline
    \textbf{MW Potential} & $\bf{\rho_0}$ & $\bf{\nu}$ \\ 
     & $\mathrm{(M_\odot\,pc^{-3})}$ &$\mathrm{(km\,s^{-1}\,kpc^{-1})}$\\\hline\hline
     MWP14-1 & 0.096 & 73.3   \\
     MWP14-2 & 0.100 & 74.2  \\
     MWP14-3 & 0.104 & 75.0  \\
     McMillan17 & 0.114 & 78.6  \\
     Irrgang13I & 0.102 & 74.8
\end{tabular}
\caption{Disc properties of the Milky Way-like potentials considered in \secname \ref{sec:MWmodels} and \ref{sec:sgrA}.}
\label{tb:pot_p0_vf}
\end{table}

For the distribution function in the model, we use a 3-component quasi-isothermal distribution function. with velocity dispersion and surface densities taken from \figurename 8 of \citet{joDF}. The three components have a surface density of 21, 5.8 and 2.44 $\mathrm{M_\odot\,pc^{-2}}$ and velocity dispersions of 23.2, 36.6, and 46.0 $\mathrm{km\,s^{-1}}$. We then use these values to find the number density for each component of the distribution function.

\figurename \ref{fig:realMWComp} shows us the asymmetry created by Sgr 2 starting two apocentres ago and ending at its position today (rather than at the next apocentre as we had been considering in \secname \ref{sec:simple_model} above). For the most part, the asymmetry exhibits the expected behaviour as discussed in \secname \ref{sec:mpd}. For four of the potentials, as the vertical frequency increases, the wavelength  of the asymmetry increases as well. However, the \texttt{McMillan17} potential which has the highest vertical frequency of all considered potentials, also has an asymmetry perturbation wavelength roughly equivalent to that of MWP14-2 and \texttt{Irrgang13I}. Upon further investigation, we find that while orbits that start near the mid-plane with a vertical velocity of approximately zero do have a larger frequency than the other potentials, as one moves away from the mid-plane or increases the velocity of the orbit, the frequency of the orbit aligns well with those from other potentials. In other words, low energy orbits have higher frequencies, but higher energy orbits have similar frequencies to the other four potentials. This is likely because the vertical profile of the \texttt{McMillan17} potential is much more complex than any of the other potentials and includes the two gas discs which have much smaller scale heights than the stellar disc in any of the other models. This tells us that while the vertical frequency of the disc is a good indicator of how a potential will behave relative to other potentials, the vertical frequency of specific points in phase-space matters even more and determining the behaviour of a more complicated disc is not altogether straight forward.

We are now using a realistic potential for the disc, a more realistic distribution function, and a realistic orbit of Sgr. This means that the asymmetry can now be compared to the observed asymmetry in \citet{ME}, shown by the black points in \figurename \ref{fig:realMWComp}. Clearly the model does not match the observed asymmetry, but there are still many more combinations of orbits, Sgr mass models, and disc potentials to consider. We will explore these combinations in \secname \ref{sec:sgrA} and discuss how they compare to the observations.

\subsection{Comparing model at present day to Observations}\label{sec:sgrA}

So far we have discussed a simple yet realistic model of the asymmetry. We have also looked at the resulting asymmetry in realistic models of the Milky Way's gravitational field and of the local stellar distribution function in \secname \ref{sec:MWmodels}. However, we have not yet fully investigated how this model could be applied to all the possible orbits and mass models of the Sagittarius dwarf galaxy. To do this, we combine the orbit investigation of \secname \ref{sec:sgr_orbit} with the realistic Milky-Way models of \secname \ref{sec:sgr_model} to investigate the full array of asymmetries that could be caused by Sgr. As mentioned in the previous section, using realistic orbits for Sagittarius and a more realistic disc potential and distribution function in our model, we are able to compare our results to the \emph{Gaia} asymmetry measured in \citet{ME} to determine whether or not we should blame Sgr for the observed perturbation to the vertical disc or not. 

We found in \secname \ref{sec:sgr_orbit} that the relationship between time since passing through the mid-plane and the velocity at which Sgr passed through the midplane is highly correlated. In this section, for each of the 15 combinations of Sgr models and Milky Way models mentioned in \secname \ref{sec:sgr_orbit}, we choose 10 representative orbits to show the asymmetries resulting from these Sgr orbits. We do this by binning along the velocity at which Sgr went through the mid-plane, $v_{z,\mathrm{through}}$, such that each bin contains 1001 orbits. From there, for each bin, we choose the orbit with the median value in $v_{\mathrm{z,\mathrm{through}}}$. We then integrate these ten orbits from present day back through two pericentres to the apocentre.   
\figurename \ref{fig:Sgr_Asym} shows the asymmetries for the different representative orbits for each model as well as the real asymmetry measured by \emph{Gaia}. Excluding some outliers, as the Sgr models run from heavier to lighter, the amplitude of the asymmetry decreases as expected. The outliers arise when the period of the satellites orbit matches the rotational period of the Sun around the centre of the Milky Way. This happens constructively to the second fastest orbit in MWP14-2 and destructively to the three fastest orbits in MWP14-3. We also see that the wavelength of the asymmetry increases as we move from faster velocities (red) to slower velocities (blue) as Sgr passes through the disc. This trend is not affected by the rotational period of the Sun like the amplitude and changes consistently across all models. Though not shown here, the median vertical velocity from the model also does not match for any combination of Milky-Way and Sgr models. The amplitude of the mean vertical velocity in the model is consistently too small relative to the observed mean velocity by an order of magnitude or more. It is also evident that none of the models match the observed asymmetry. The most promising of the models is the second fastest Sgr passage (dark orange) with Sgr mass model 4 and Milky Way potential MWP14-3. However, while the perturbation wavelength matches within 0.7 kpc of the mid-plane, it does not match further out. The amplitude also does not match the observed asymmetries. Finally, when we look at the mean vertical velocity of the model, both the amplitude and the perturbation wavelength don't match the observations. We will discuss why changing parameters of the disc could not ameliorate the match between observations and the model in \secname \ref{sec:match}.

In \secname \ref{sec:peri}, we investigated the effect of multiple pericentres of the satellite on the perturbation and found that the most recent pericentre sets the overall shape of the asymmetry while the additional pericentres add smaller scale fluctuations to the asymmetry. Based on this, we can almost certainly rule out a mass of $5.1\times10^{10}$ or $1.02\times 10^{10}\,\mathrm{M_\odot}$ for Sgr on its most recent pericentre passage as the amplitudes of the resultant asymmetries are much too large. This is not in conflict with \citet{Laporte2019}, which considered Sgr's mass approximately $5-6$ Gyr ago and does not discuss the mass at the most recent pericentre passage. However, it does disagree with \citet{BH-GALAH}, which predicts the mass at the most recent pericentre passage to be $5\times10^10\,\mathrm{M_\odot}$ stripped down to $3\times 10^{10}\,\mathrm{M_\odot}$. Given that the present-day mass of the Sgr progenitor was recently measured to be $1-4\times 10^8\,\mathrm{M_\odot}$ \citep{SgrModel}, we suspect that the discrepancy between the mass estimates in \citet{BH-GALAH} is likely due to the fact that all these models assume Sgr was the sole cause of the perturbation, but we have shown that that cannot be the case. In the following section we discuss what it would take to replicate the observed asymmetry by changing the properties of the Milky Way disc.

\subsection{Matching observations} \label{sec:match}

\begin{figure}
    \centering
    \includegraphics[width=0.45\textwidth]{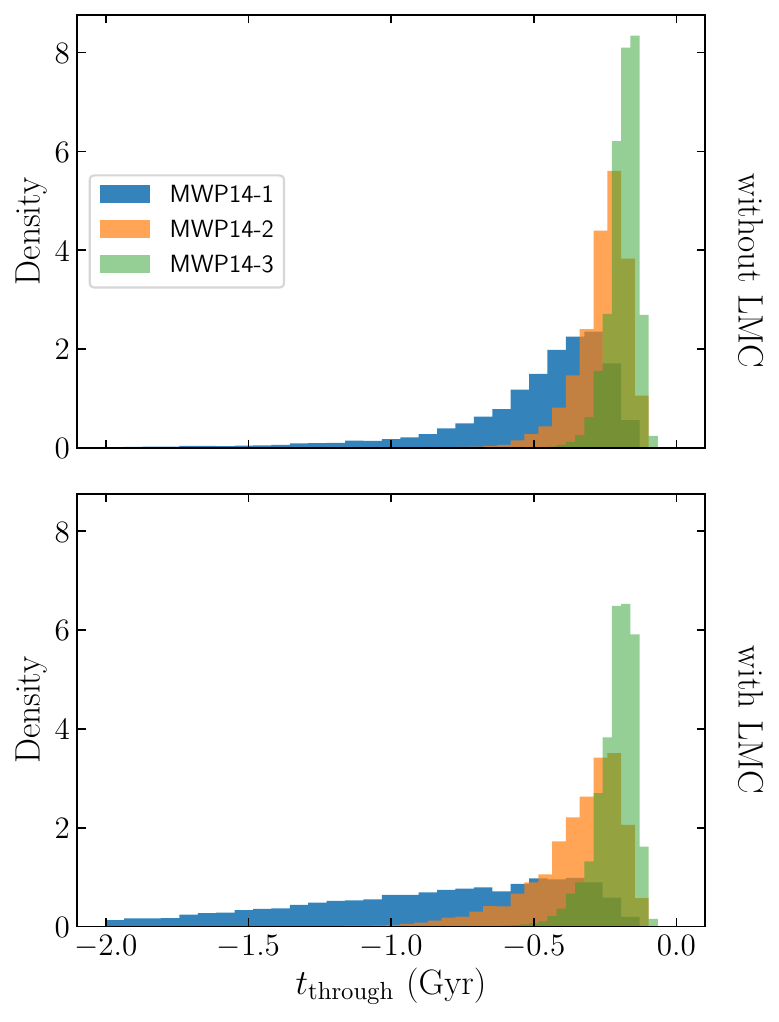}
    \caption{Comparison of impact property of Sgr's last passage through the disc with and without the effects if the LMC. \textit{Top:} Distribution of the time at which Sagittarius passed through the mid-plane of the disc for the middle Sagittarius mass model, Sgr 3, in each of the Milky Way models.
    \textit{Bottom:} Distribution of $t_{\mathrm{through}}$, for the same Sgr model when integrated with a moving object potential corresponding to the properties of the LMC.  
    The addition of the LMC has caused the distributions to widen and shift to earlier times, corresponding to Sgr passing through the disc longer ago.}
    \label{fig:tthrough_LMC}
\end{figure}

\begin{figure*}
    \centering
    \includegraphics[width=0.9\textwidth]{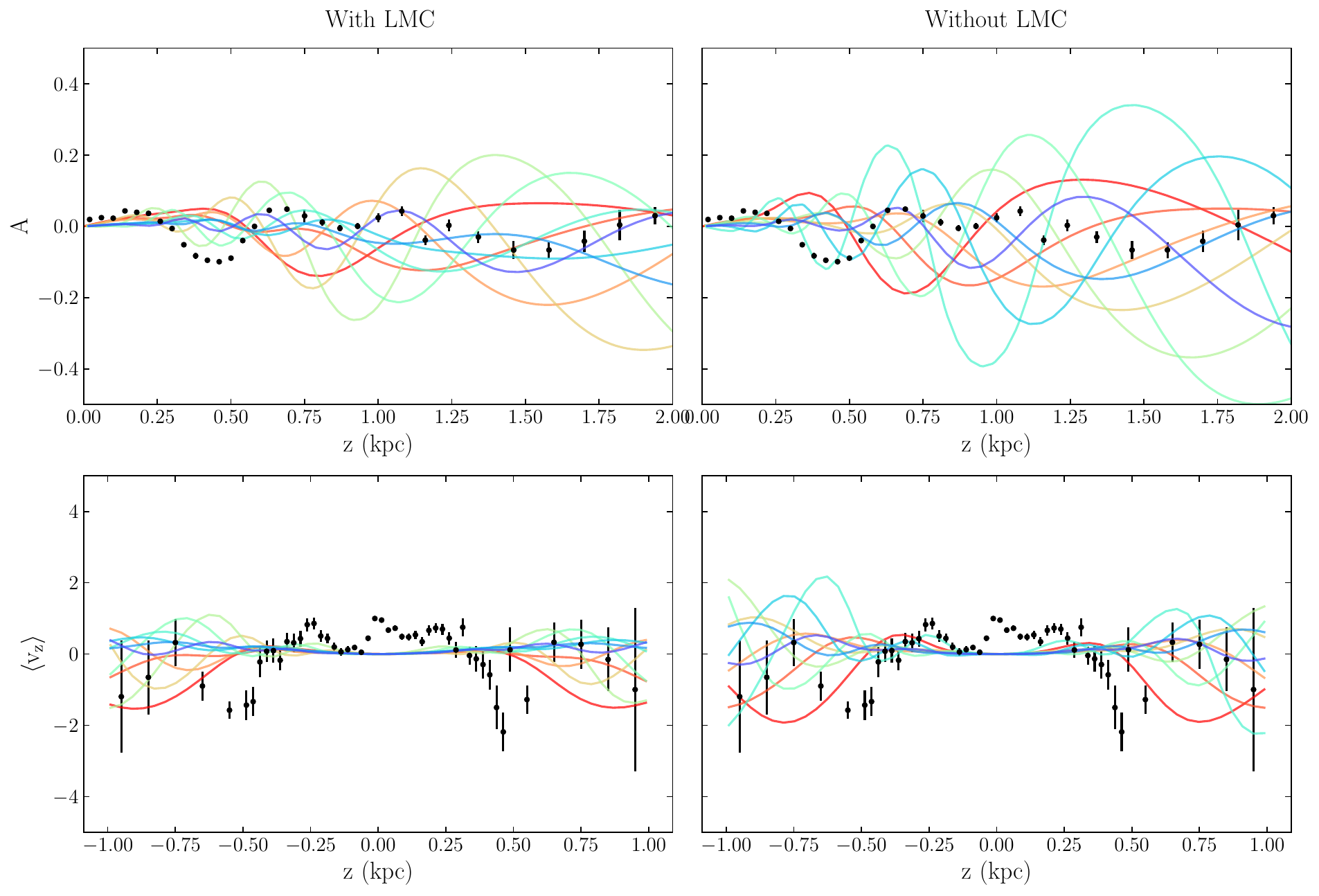}
    \caption{Perturbation from Sgr 3 in MWP14-3 for the case without the LMC (corresponding to the bottom middle panel of \figurename \ref{fig:Sgr_Asym}) (left) and the case with the LMC (right). The top row shows the asymmetry, and the bottom row shows the mean vertical velocity for each model. Each plot has the model calculated for 10 representative orbits ranging from fastest through the mid-plane (red) to slowest through the mid-plane (blue). The black points in each plot are the observations from \emph{Gaia} DR2 taken from \citet{ME}. None of the models reproduce the features of both the asymmetry and the mean vertical velocity.}
    \label{fig:bestA}
\end{figure*}

\begin{figure}
    \centering
    \includegraphics[width=0.45\textwidth]{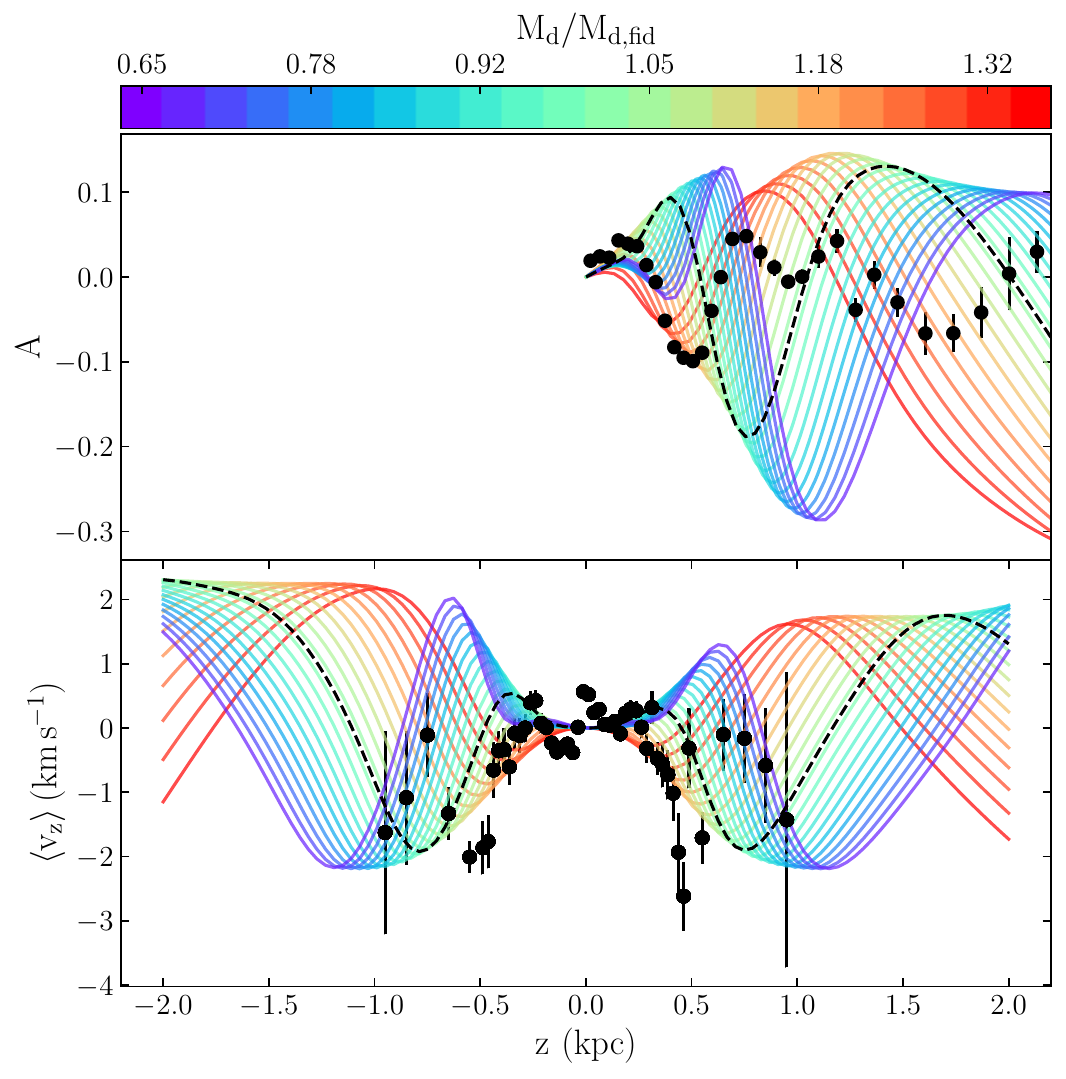}
    \caption{Perturbation given a range of lighter (blue) to heavier (red) discs for the fastest Sgr orbit using the heaviest Milky Way potential, MWP14-3, and the middle Sagittarius mass model, Sgr3. $\mathrm{M_d/M_{d,fid}}$ is the ratio between the model disc mass and the disc mass of the fiducial model, MWP14-3. We consider the fastest orbits from this combination of models, because it is the closest visual match to the measured asymmetry. The perturbation for that model with no adjustments to the disc mass is also plotted on both the top and bottom panel (black dashed line). Top: Number count asymmetry for the varying disc masses as well as the observed asymmetry (black points). Bottom: Mean vertical velocity for the same models and observed mean vertical velocity (black points). While the change to the disc mass does make some modelled perturbations more closely resemble data, none are a match.}
    \label{fig:matchA_LMC}
\end{figure}

In the previous section, we discovered that the perturbation due to Sgr does not match the observed asymmetry for any of the possible Sgr orbits, Sgr mass models, or Milky Way potentials. Additionally, we discussed that the vertical mean velocity does not agree with observations for any of the models. In this section we will discuss if there are any alterations we could make to the properties of our model that would change the asymmetry enough to make it match the observed \emph{Gaia} DR2 data. We look at the effects of the Large Magellenic Clouds (LMC) on the orbit of Sgr, and then see if we can tweak the properties of the disc to make the model match observations. 

To investigate the effects the LMC has on the orbit of Sgr, we use the same method as discussed in \secname \ref{sec:sgr_orbit} to integrate Sgr's orbit back in time, but also include a \texttt{galpy} moving object potential for the LMC which implements a potential along the LMC's orbit during the integration of Sgr's orbit. We use the parameters discussed in \citet{Vasiliev2020}. For the potential of the LMC, we use a Hernquist potential with a mass of $2.\times10^{11}\,\mathrm{M_\odot}$ and a scale radius of $12.9\,\mathrm{kpc}$. We chose the heaviest of their LMC models as it will have the largest effect on our model. To initialise the orbit, we use the same current day position and velocities, $x,y,z = \{-0.6,\,-41.3,\, -27.1\}\,\,\mathrm{kpc}$, and $v_x,v_y,v_z =
\{-63.9,\,-213.8,\,206.6\}\,\,\mathrm{km\,s^{-1}}$. We do not consider the error in the LMC parameters as the effects from the Sgr orbit uncertainties and uncertainty in the halo mass are already quite substantial. Overall, we find that with the inclusion of the LMC, the median time since impact increases for each of our Sgr and Milky Way models. This is shown in \figurename \ref{fig:tthrough_LMC} where we compare the distribution of the time since passing through with and without the LMC for MWP14-3 and Sgr 3 and find that in all cases the distribution has widened and shifted to earlier times. Though not shown here, this trend is also evident in the other combinations of Sgr mass models and Milky Way potential models. As seen in \secname \ref{sec:real}, an earlier crossing time means that the asymmetry has a smaller wavelength. Therefore, the inclusion of the LMC shortens the perturbation wavelength as more time has passed since the perturbation resulting in more phase-mixing. In most cases, the asymmetry wavelength is too short, so for most models this means that adding the LMC makes the models and observations disagree even more drastically. 

As we showed in \secname \ref{sec:MWmodels}, changing the disc model does not have a large impact on the overall shape of the asymmetry. It can change the wavelength and amplitude of the asymmetry, but it appears the orbit of the satellite is what determines the shape of the asymmetry. This means that if we want to adjust the parameters of the disc to make it match the observations, we have to start with an asymmetry and mean vertical velocity which has a shape similar to the data, though the perturbation wavelength and period can be off. For this reason, and because the addition of the LMC shortens the wavelength, we choose to look at the perturbation models from Sgr 3 and MWP14-3, which have an amplitude similar to the observations and a shape which looks approximately correct, though the asymmetry wavelength is long compared to the observed asymmetry.

The first thing to consider when trying to make the models match the observation is whether or not the addition of self-gravity could bridge the gap between the model and the observations. As we saw in \secname \ref{sec:alpha}, both the amplitude and the perturbation wavelength increases as we increase $\alpha$ and therefore the self-gravity of the simulations. It also shifts the peaks inwards. None of these changes would help lessen the discrepancy between the model and observations, as the wavelength of the asymmetry is already too long in the model and in most cases, the amplitude of the model is already consistent with or larger than the observations. This tells us that self-gravity does not alleviate the discrepancy between the models and the data, so we must look at other methods of making the two agree.

Next we look at how changing our disc potential and distribution function could match the model to the observations. The most promising of the models is the fastest Sgr 3 in MWP14-3 from \figurename \ref{fig:bestA} which resemble the observed asymmetry in shape, if not the perturbation wavelength. In \secname \ref{sec:mpd}, we saw that changing the mid-plane density of the potential can change the wavelength of the asymmetry. \secname \ref{sec:MWmodels} showed us that while this assumption becomes more complicated as you increase the complexity of the Milky Way potential, it does still hold for the different MWP14s. For that reason, we introduce a multiplicative factor to the disc mass ranging between 0.65 and 1.35 which has the effect of changing the mid-plane density and allows us to explore both lighter and heavier discs for the same Sgr orbit. \figurename \ref{fig:matchA_LMC} shows the models for each of these mid-plane densities. As mentioned, we explored discs which range between 0.65 of the mass of MWP14-3's disc and 1.35 the disc mass. None of these models reproduce the wavelength of the asymmetry and the mean vertical velocity. With a 25\% shift downwards in the disc mass, the model matches the first dip in the asymmetry. To match the dips in the mean vertical velocity requires a downward shift of 19\% in the disc mass. However, none of the models manage to match the observed asymmetry beyond the first dip.

Another limitation of our model is that it is purely one dimensional and therefore does not capture the range of vertical frequencies that are present  in the solar neighbourhood at a given vertical action due to the vertical orbital frequency's dependence on, e.g., the angular momentum (which is ultimately the reason that the \emph{Gaia} phase-space spiral stands out so strongly when colour-coding the vertical phase space by angular momentum; \citealt{Binney2018}). Using a realistic three-dimensional quasi-isothermal distribution function \citep{binney2011} for an old stellar population in the solar neighbourhood and computing the vertical actions and frequencies for a sample of stars sampled from it in \texttt{MWPotential2014}, we determine that the spread in vertical frequency at a given vertical action is about 20\,\%, with an overall mean shift of about 10\,\% (due to the asymmetric drift). These frequency differences from the one-dimensional model are smaller than the $\sim 25\,\%$ shift in mass required above to get a better match to the observed asymmetry and velocity, and as we discussed before, even with this shift, a good match cannot really be said to have been attained as the velocity and asymmetry never agree with the data at the same time. 

We have established that adding self-gravity to the disc does not help with reproducing the observed asymmetry in the disc for any Sgr model. We have also discussed the parameters of the model we would have to tweak to make it agree with observations. Even when we consider the effects of the LMC, it does not change Sgr's orbit enough to find a match. Not only did we have to choose the fastest moving Sgr orbit, but we also had to change the mass of the disc by $\sim 25\%$. This is significantly outside the margin of error measured for the total mid-plane density of the disc \citep[e.g.,][]{p0-2} and we also argued that the spread in vertical frequencies coming from the spread in angular momenta in the solar neighbourhood (in a full three-dimensional disc model) is too small to significantly affect our conclusions. After making all these changes, we still could not match both the observed asymmetry and the mean vertical velocity. These all stack up against Sgr being the cause of the observed perturbation to the vertical disc.

\section{Conclusion}\label{sec:conclusion}

The discovery of the phase-space spiral in the solar neighbourhood in \emph{Gaia} DR2 has brought about much excitement in the world of galactic dynamics. Many investigations into the properties and the cause of the spiral have been investigated over the past few years. Among the foremost of the origin theories is the idea that the passing of Sgr perturbed the disc and led to the oscillations we see today. Throughout this paper, we explored whether or not this could truly be the case.

We started by introducing a new method for calculating the one-dimensional vertical perturbation to the Galactic disc given a vertical force. We use Liouville's theorem to relate the perturbed distribution function today to an unperturbed distribution function at some initial time in the past. Next, we calculate the change in the action of an orbit in the disc given a perturbing force. We use that to calculate the perturbed distribution function which can be integrated over $v_z$ to find the perturbed density of the disc or can be used to take the first moment of $v_z$ to calculate the mean vertical velocity as a function of height. While our formalism works for an arbitrary force, in this paper we choose to focus on the force from a passing satellite.

In \secname \ref{sec:simple_model}, we looked at a simple example of the model given the force from a satellite that goes through one orbit from apocentre to apocentre. Using this simple set-up, we looked at some properties of our model and how the changing the different components effects the results. First, we considered how changing the mid-plane density of the disc changes the asymmetry and mean vertical velocity. We found that as we increase the mid-plane density, and therefore the mass and vertical frequency of our disc, we also increase the amplitude and perturbation wavelength of the oscillations in the asymmetry and mean vertical velocity. Next, we looked at how considering different tracers of the disc affects our model by varying the velocity dispersion of the distribution function within a fixed gravitational potential. We chose velocity dispersions ranging between 10 and 40 $\mathrm{km\,s^{-1}}$ to reflect the range of observed velocity dispersions in the solar neighbourhood. As we increase the velocity dispersion, the amplitude of the asymmetry and mean vertical velocity decreases significantly, but the wavelength  of the asymmetry does not noticeably change. Finally, we looked at how the number of pericentre passages of the satellite affects the observed asymmetry. The difference between one and two pericentre passages was minor within 1 kpc, but quite large further out. The difference between two and three pericentre passages was not significant within 2 kpc. We determined that going back two pericentres was the best balance between accuracy in the orbit and capturing behaviours of the perturbation. 

Next, we compared our model to a one-dimensional simulation with varying amounts of self-gravity. With minimal self-gravity, the non-linear perturbation in the simulation agreed well with the linear perturbation calculation of our model. We then compared the asymmetry as a function of time and $z$ in the model to simulations with increasing self-gravity. We found that as we increase self-gravity, the amplitude and wavelength  of the asymmetry increased. We also found that the shape of the asymmetry began to be shifted inwards in $z$ as we increased self-gravity and the oscillations became washed out. We also noted that as we increase the self-gravity in the simulation, the disc begins to behave like a rigid body and is dragged around by the satellite passage. For the case where the disc is fully self-gravitating, the disc was dragged around, but very few internal perturbations arose. However, a fully self-gravitating disc is unrealistic for a one-dimensional simulation because the stars at different radii, as well as the dark matter distribution, are better modeled as a fixed potential on the solar neighbourhood. 

Finally, in \secname \ref{sec:real}, we consider a more realistic scenario for our model using Sgr and comparing it to the asymmetry and mean vertical velocity observed by \emph{Gaia} DR2. We start by discussing the different mass models for Sgr, which range between $5.1\times 10^{10}-5.1\times 10^{8}\,\mathrm{M_\odot}$ and include both a stellar and dark matter component represented by Hernquist potentials. Next, we looked at how these five different Sgr mass models orbit in 3 different Milky Way potentials derived from \texttt{MWPotential2014} and with increasing halo mass. We considered two properties of the Sgr orbits, the time since Sgr passed through the disc and the speed at which it was travelling when it passed through the disc. We found that the Milky Way potential has a much bigger impact on these two variables than the Sgr mass model does. 

After investigating the properties of Sgr, we considered how five different realistic Milky Way potentials would behave given the same perturbing force. We found that our simplistic view of the relation between the vertical frequency and the perturbation wavelength of the oscillation due to the perturbation is actually more complicated. When comparing the same potential scaled to different masses, the relation holds, but comparing two potentials with different shapes complicates matters and we can no longer predict the relative wavelength  of the asymmetry for the two potentials based on their vertical frequencies. This tells us that the perturbation wavelength depends more on the period of the orbit at a single point in phase-space than it does on the overall vertical frequency of the disc.

Having developed and thoroughly tested the fast one-dimensional model for the perturbation from a satellite and having set up realistic models of the Milky Way's disc populations and gravitational potential, we turn to the main question asked in the title of this paper: Did Sgr cause the vertical waves in the solar neighbourhood? To answer this, we looked at ten representative orbits of Sgr for each combination of the five Sgr mass models and three Milky Way models. We chose these by binning the orbits by the speed at which they went through the mid-plane of the disc, and then taking the median orbit from each of the ten bins---these median orbits span the range of Sgr orbit properties relevant for calculating the vertical perturbation. For each orbit, with each Sgr mass model and each Milky Way model, we calculated the perturbation to the vertical disc and compared it to the observed asymmetry and mean vertical velocity observed in \emph{Gaia} DR2. We found that none of the models even remotely matched the data in the asymmetry or in the mean vertical velocity. The few that were similar in shape and amplitude were still off by a factor of $\approx 2$ in the perturbation wavelength. This is the same as the findings of \citet{Laporte2019} in their N-body simulation. From the asymmetries predicted by our model, we also suggest that the mass of Sgr on its most recent pericentre passage must have been below $10^{10}\,\mathrm{M_\odot}$. Finally, we looked at how we could modify the model to try and better match the observations. We started by considering the effect of the LMC on Sgr's orbit and found that it consistently makes Sgr pass through the disc earlier, resulting in a shorter asymmetry wavelength. We next considered how we could change the properties of the disc to match the \emph{Gaia} DR2 observations by looking at a range of disc masses and found that when we changed the mass of the disc by 25\% we were able to match the first feature of the asymmetry but not the mean vertical velocity and a shift of 19\% made the mean vertical velocity match, but left the asymmetry in great disagreement with observations. No one model was able to match both. Which means that even with these extreme measures, we could not make the perturbation due to Sgr match the observed asymmetry and mean vertical velocity of the disc.

While we find that satellites are capable of generating phase-space spirals and therefore asymmetries and mean vertical velocities qualitatively similar to those observed in Gaia DR2, the Sagittarius dwarf galaxy is not able to reproduce observations. Even making unrealistic modifications to the disc could not resolve the difference between the model and the data. Given that the progenitor mass of Sgr was recently measured to be less than $5\times10^{10}\,\mathrm{M_\odot}$, our model showed only one of thirty cases where the asymmetry signal due to Sgr would be above the level of errors in the observations. This tells us that Sgr could not have caused the perturbation to the disc seen in phase-space and the velocity field. It is more likely that the true perturbation from Sgr is one of the scenarios where the amplitude is small and therefore contributes minimally to the observed asymmetry. It is possible that the perturbation of the disc was caused by the passage of another satellite or some internal factor, although that has yet to be determined. All that can be said for certain is that Sgr is not the main contributor to the vertical perturbation in the solar neighbourhood.

%%%%%%%%%%%%%%%%%%%%%%%%%%%%%%%%%%%%%%%%%%%%%%%%%%

\section*{Acknowledgements}

It is a pleasure to thank Jason Hunt and Daisuke Kawata for helpful discussions. MB and JB acknowledge financial support from NSERC (funding reference numbers RGPIN-2015-05235 \& RGPIN-2020-04712) and an Ontario Early Researcher Award (ER16-12-061). This work has made use of data from the European Space Agency (ESA) mission
{\it Gaia} (\url{https://www.cosmos.esa.int/gaia}), processed by the {\it Gaia}
Data Processing and Analysis Consortium (DPAC,
\url{https://www.cosmos.esa.int/web/gaia/dpac/consortium}). Funding for the DPAC
has been provided by national institutions, in particular the institutions
participating in the {\it Gaia} Multilateral Agreement.

\section{Data availability statement}
The data underlying this article are available in the article and in its online supplementary material found at \url{https://github.com/morganb-phys/Sgr-VerticalWaves}.



\bibliographystyle{mnras}
\bibliography{references} % if your bibtex file is called example.bib [but give it a better name!]

\begin{thebibliography}{}
\makeatletter
\relax
\def\mn@urlcharsother{\let\do\@makeother \do\$\do\&\do\#\do\^\do\_\do\%\do\~}
\def\mn@doi{\begingroup\mn@urlcharsother \@ifnextchar [ {\mn@doi@}
  {\mn@doi@[]}}
\def\mn@doi@[#1]#2{\def\@tempa{#1}\ifx\@tempa\@empty \href
  {http://dx.doi.org/#2} {doi:#2}\else \href {http://dx.doi.org/#2} {#1}\fi
  \endgroup}
\def\mn@eprint#1#2{\mn@eprint@#1:#2::\@nil}
\def\mn@eprint@arXiv#1{\href {http://arxiv.org/abs/#1} {{\tt arXiv:#1}}}
\def\mn@eprint@dblp#1{\href {http://dblp.uni-trier.de/rec/bibtex/#1.xml}
  {dblp:#1}}
\def\mn@eprint@#1:#2:#3:#4\@nil{\def\@tempa {#1}\def\@tempb {#2}\def\@tempc
  {#3}\ifx \@tempc \@empty \let \@tempc \@tempb \let \@tempb \@tempa \fi \ifx
  \@tempb \@empty \def\@tempb {arXiv}\fi \@ifundefined
  {mn@eprint@\@tempb}{\@tempb:\@tempc}{\expandafter \expandafter \csname
  mn@eprint@\@tempb\endcsname \expandafter{\@tempc}}}

\bibitem[\protect\citeauthoryear{{Allen} \& {Santillan}}{{Allen} \&
  {Santillan}}{1991}]{IrrgangHalo}
{Allen} C.,  {Santillan} A.,  1991, \rmxaa, \href
  {https://ui.adsabs.harvard.edu/abs/1991RMxAA..22..255A} {22, 255}

\bibitem[\protect\citeauthoryear{{Antoja} et~al.,}{{Antoja}
  et~al.}{2018}]{antoja}
{Antoja} T.,  et~al., 2018, \mn@doi [Nature] {10.1038/s41586-018-0510-7}, \href
  {https://ui.adsabs.harvard.edu/abs/2018Natur.561..360A} {561, 360}

\bibitem[\protect\citeauthoryear{{Bennett} \& {Bovy}}{{Bennett} \&
  {Bovy}}{2019}]{ME}
{Bennett} M.,  {Bovy} J.,  2019, \mn@doi [\mnras] {10.1093/mnras/sty2813},
  \href {https://ui.adsabs.harvard.edu/abs/2019MNRAS.482.1417B} {482, 1417}

\bibitem[\protect\citeauthoryear{Binney}{Binney}{2010}]{binney10}
Binney J.,  2010, \mn@doi [\mnras] {10.1111/j.1365-2966.2009.15845.x}, 401,
  2318

\bibitem[\protect\citeauthoryear{{Binney} \& {McMillan}}{{Binney} \&
  {McMillan}}{2011}]{binney2011}
{Binney} J.,  {McMillan} P.,  2011, \mn@doi [\mnras]
  {10.1111/j.1365-2966.2011.18268.x}, \href
  {https://ui.adsabs.harvard.edu/abs/2011MNRAS.413.1889B} {413, 1889}

\bibitem[\protect\citeauthoryear{{Binney} \& {Sch{\"o}nrich}}{{Binney} \&
  {Sch{\"o}nrich}}{2018}]{Binney2018}
{Binney} J.,  {Sch{\"o}nrich} R.,  2018, \mn@doi [\mnras]
  {10.1093/mnras/sty2378}, \href
  {https://ui.adsabs.harvard.edu/abs/2018MNRAS.481.1501B} {481, 1501}

\bibitem[\protect\citeauthoryear{{Binney} \& {Tremaine}}{{Binney} \&
  {Tremaine}}{2008}]{binney&tremaine}
{Binney} J.,  {Tremaine} S.,  2008, {Galactic Dynamics: Second Edition}

\bibitem[\protect\citeauthoryear{{Bissantz} \& {Gerhard}}{{Bissantz} \&
  {Gerhard}}{2002}]{McmillanBulge}
{Bissantz} N.,  {Gerhard} O.,  2002, \mn@doi [\mnras]
  {10.1046/j.1365-8711.2002.05116.x}, \href
  {https://ui.adsabs.harvard.edu/abs/2002MNRAS.330..591B} {330, 591}

\bibitem[\protect\citeauthoryear{{Bland-Hawthorn} et~al.,}{{Bland-Hawthorn}
  et~al.}{2019}]{BH-GALAH}
{Bland-Hawthorn} J.,  et~al., 2019, \mn@doi [\mnras] {10.1093/mnras/stz217},
  \href {https://ui.adsabs.harvard.edu/abs/2019MNRAS.486.1167B} {486, 1167}

\bibitem[\protect\citeauthoryear{{Bovy}}{{Bovy}}{2015}]{galpy}
{Bovy} J.,  2015, \mn@doi [\apjs] {10.1088/0067-0049/216/2/29}, \href
  {https://ui.adsabs.harvard.edu/abs/2015ApJS..216...29B} {216, 29}

\bibitem[\protect\citeauthoryear{{Bovy} \& {Rix}}{{Bovy} \&
  {Rix}}{2013}]{BovyRix}
{Bovy} J.,  {Rix} H.-W.,  2013, \mn@doi [\apj] {10.1088/0004-637X/779/2/115},
  \href {https://ui.adsabs.harvard.edu/abs/2013ApJ...779..115B} {779, 115}

\bibitem[\protect\citeauthoryear{{Bovy}, {Rix}, {Hogg}, {Beers}, {Lee}  \&
  {Zhang}}{{Bovy} et~al.}{2012}]{joDF}
{Bovy} J.,  {Rix} H.-W.,  {Hogg} D.~W.,  {Beers} T.~C.,  {Lee} Y.~S.,   {Zhang}
  L.,  2012, \mn@doi [\apj] {10.1088/0004-637X/755/2/115}, \href
  {https://ui.adsabs.harvard.edu/abs/2012ApJ...755..115B} {755, 115}

\bibitem[\protect\citeauthoryear{{Carrillo} et~al.,}{{Carrillo}
  et~al.}{2019}]{Carillo}
{Carrillo} I.,  et~al., 2019, \mn@doi [\mnras] {10.1093/mnras/stz2343}, \href
  {https://ui.adsabs.harvard.edu/abs/2019MNRAS.490..797C} {490, 797}

\bibitem[\protect\citeauthoryear{{Chequers}, {Widrow}  \& {Darling}}{{Chequers}
  et~al.}{2018}]{Chequers2018}
{Chequers} M.~H.,  {Widrow} L.~M.,   {Darling} K.,  2018, \mn@doi [\mnras]
  {10.1093/mnras/sty2114}, \href
  {https://ui.adsabs.harvard.edu/abs/2018MNRAS.480.4244C} {480, 4244}

\bibitem[\protect\citeauthoryear{{Darling} \& {Widrow}}{{Darling} \&
  {Widrow}}{2019}]{widrow_alpha}
{Darling} K.,  {Widrow} L.~M.,  2019, \mn@doi [\mnras] {10.1093/mnras/stz2539},
  \href {https://ui.adsabs.harvard.edu/abs/2019MNRAS.490..114D} {490, 114}

\bibitem[\protect\citeauthoryear{{Faure}, {Siebert}  \& {Famaey}}{{Faure}
  et~al.}{2014}]{Faure2014}
{Faure} C.,  {Siebert} A.,   {Famaey} B.,  2014, \mn@doi [\mnras]
  {10.1093/mnras/stu428}, \href
  {https://ui.adsabs.harvard.edu/abs/2014MNRAS.440.2564F} {440, 2564}

\bibitem[\protect\citeauthoryear{{Gaia Collaboration} et~al.,}{{Gaia
  Collaboration} et~al.}{2018a}]{Gaia}
{Gaia Collaboration} et~al., 2018a, \mn@doi [\aap]
  {10.1051/0004-6361/201833051}, \href
  {https://ui.adsabs.harvard.edu/abs/2018A&A...616A...1G} {616, A1}

\bibitem[\protect\citeauthoryear{{Gaia Collaboration} et~al.,}{{Gaia
  Collaboration} et~al.}{2018b}]{GaiaKinematics}
{Gaia Collaboration} et~al., 2018b, \mn@doi [\aap]
  {10.1051/0004-6361/201832865}, \href
  {https://ui.adsabs.harvard.edu/abs/2018A&A...616A..11G} {616, A11}

\bibitem[\protect\citeauthoryear{{Gaia Collaboration} et~al.,}{{Gaia
  Collaboration} et~al.}{2018c}]{GaiaSatKin}
{Gaia Collaboration} et~al., 2018c, \mn@doi [\aap]
  {10.1051/0004-6361/201832698}, \href
  {https://ui.adsabs.harvard.edu/abs/2018A&A...616A..12G} {616, A12}

\bibitem[\protect\citeauthoryear{{G{\'o}mez}, {Minchev}, {O'Shea}, {Beers},
  {Bullock}  \& {Purcell}}{{G{\'o}mez} et~al.}{2013}]{Gomez2013}
{G{\'o}mez} F.~A.,  {Minchev} I.,  {O'Shea} B.~W.,  {Beers} T.~C.,  {Bullock}
  J.~S.,   {Purcell} C.~W.,  2013, \mn@doi [\mnras] {10.1093/mnras/sts327},
  \href {https://ui.adsabs.harvard.edu/abs/2013MNRAS.429..159G} {429, 159}

\bibitem[\protect\citeauthoryear{{Gravity Collaboration} et~al.,}{{Gravity
  Collaboration} et~al.}{2019}]{Gravity}
{Gravity Collaboration} et~al., 2019, \mn@doi [\aap]
  {10.1051/0004-6361/201935656}, \href
  {https://ui.adsabs.harvard.edu/abs/2019A&A...625L..10G} {625, L10}

\bibitem[\protect\citeauthoryear{{Hernquist}}{{Hernquist}}{1990}]{Hernquist}
{Hernquist} L.,  1990, \mn@doi [\apj] {10.1086/168845}, \href
  {https://ui.adsabs.harvard.edu/abs/1990ApJ...356..359H} {356, 359}

\bibitem[\protect\citeauthoryear{{Holmberg} \& {Flynn}}{{Holmberg} \&
  {Flynn}}{2000}]{p0-1}
{Holmberg} J.,  {Flynn} C.,  2000, \mn@doi [\mnras]
  {10.1046/j.1365-8711.2000.02905.x}, \href
  {https://ui.adsabs.harvard.edu/abs/2000MNRAS.313..209H} {313, 209}

\bibitem[\protect\citeauthoryear{{Hunt}, {Hong}, {Bovy}, {Kawata}  \&
  {Grand}}{{Hunt} et~al.}{2018}]{Hunt2019}
{Hunt} J. A.~S.,  {Hong} J.,  {Bovy} J.,  {Kawata} D.,   {Grand} R. J.~J.,
  2018, \mn@doi [\mnras] {10.1093/mnras/sty2532}, \href
  {https://ui.adsabs.harvard.edu/abs/2018MNRAS.481.3794H} {481, 3794}

\bibitem[\protect\citeauthoryear{{Irrgang}, {Wilcox}, {Tucker}  \&
  {Schiefelbein}}{{Irrgang} et~al.}{2013}]{Irrgang13I}
{Irrgang} A.,  {Wilcox} B.,  {Tucker} E.,   {Schiefelbein} L.,  2013, \mn@doi
  [\aap] {10.1051/0004-6361/201220540}, \href
  {https://ui.adsabs.harvard.edu/abs/2013A&A...549A.137I} {549, A137}

\bibitem[\protect\citeauthoryear{{Kalnajs}}{{Kalnajs}}{1973}]{kalnajs72}
{Kalnajs} A.~J.,  1973, \mn@doi [\apj] {10.1086/152023}, \href
  {http://adsabs.harvard.edu/abs/1973ApJ...180.1023K} {180, 1023}

\bibitem[\protect\citeauthoryear{{Khoperskov}, {Di Matteo}, {Gerhard}, {Katz},
  {Haywood}, {Combes}, {Berczik}  \& {Gomez}}{{Khoperskov}
  et~al.}{2019}]{Khoperskov2019}
{Khoperskov} S.,  {Di Matteo} P.,  {Gerhard} O.,  {Katz} D.,  {Haywood} M.,
  {Combes} F.,  {Berczik} P.,   {Gomez} A.,  2019, \mn@doi [\aap]
  {10.1051/0004-6361/201834707}, \href
  {https://ui.adsabs.harvard.edu/abs/2019A&A...622L...6K} {622, L6}

\bibitem[\protect\citeauthoryear{Lam, Pitrou  \& Seibert}{Lam
  et~al.}{2015}]{numba}
Lam S.~K.,  Pitrou A.,   Seibert S.,  2015, in Proceedings of the Second
  Workshop on the LLVM Compiler Infrastructure in HPC. LLVM '15.
Association for Computing Machinery, New York, NY, USA,
  \mn@doi{10.1145/2833157.2833162}

\bibitem[\protect\citeauthoryear{{Laporte}, {Johnston}, {G{\'o}mez},
  {Garavito-Camargo}  \& {Besla}}{{Laporte} et~al.}{2018}]{Laporte2018}
{Laporte} C. F.~P.,  {Johnston} K.~V.,  {G{\'o}mez} F.~A.,  {Garavito-Camargo}
  N.,   {Besla} G.,  2018, \mn@doi [\mnras] {10.1093/mnras/sty1574}, \href
  {https://ui.adsabs.harvard.edu/abs/2018MNRAS.481..286L} {481, 286}

\bibitem[\protect\citeauthoryear{{Laporte}, {Minchev}, {Johnston}  \&
  {G{\'o}mez}}{{Laporte} et~al.}{2019}]{Laporte2019}
{Laporte} C. F.~P.,  {Minchev} I.,  {Johnston} K.~V.,   {G{\'o}mez} F.~A.,
  2019, \mn@doi [\mnras] {10.1093/mnras/stz583}, \href
  {https://ui.adsabs.harvard.edu/abs/2019MNRAS.485.3134L} {485, 3134}

\bibitem[\protect\citeauthoryear{{Mackereth} et~al.,}{{Mackereth}
  et~al.}{2019}]{ted}
{Mackereth} J.~T.,  et~al., 2019, \mn@doi [\mnras] {10.1093/mnras/stz1521},
  \href {https://ui.adsabs.harvard.edu/abs/2019MNRAS.489..176M} {489, 176}

\bibitem[\protect\citeauthoryear{{McConnachie}}{{McConnachie}}{2012}]{SgrDist}
{McConnachie} A.~W.,  2012, \mn@doi [\aj] {10.1088/0004-6256/144/1/4}, \href
  {https://ui.adsabs.harvard.edu/abs/2012AJ....144....4M} {144, 4}

\bibitem[\protect\citeauthoryear{{McMillan}}{{McMillan}}{2017}]{McMillan17}
{McMillan} P.~J.,  2017, \mn@doi [\mnras] {10.1093/mnras/stw2759}, \href
  {https://ui.adsabs.harvard.edu/abs/2017MNRAS.465...76M} {465, 76}

\bibitem[\protect\citeauthoryear{{Miyamoto} \& {Nagai}}{{Miyamoto} \&
  {Nagai}}{1975}]{MN_model}
{Miyamoto} M.,  {Nagai} R.,  1975, \pasj, \href
  {https://ui.adsabs.harvard.edu/abs/1975PASJ...27..533M} {27, 533}

\bibitem[\protect\citeauthoryear{{Monari}, {Famaey}  \& {Siebert}}{{Monari}
  et~al.}{2015}]{Monari2015}
{Monari} G.,  {Famaey} B.,   {Siebert} A.,  2015, \mn@doi [\mnras]
  {10.1093/mnras/stv1206}, \href
  {https://ui.adsabs.harvard.edu/abs/2015MNRAS.452..747M} {452, 747}

\bibitem[\protect\citeauthoryear{{Monari}, {Famaey}, {Siebert}, {Grand },
  {Kawata}  \& {Boily}}{{Monari} et~al.}{2016}]{Monari2016}
{Monari} G.,  {Famaey} B.,  {Siebert} A.,  {Grand } R. J.~J.,  {Kawata} D.,
  {Boily} C.,  2016, \mn@doi [\mnras] {10.1093/mnras/stw1564}, \href
  {https://ui.adsabs.harvard.edu/abs/2016MNRAS.461.3835M} {461, 3835}

\bibitem[\protect\citeauthoryear{{Navarro}, {Frenk}  \& {White}}{{Navarro}
  et~al.}{1996}]{NFW}
{Navarro} J.~F.,  {Frenk} C.~S.,   {White} S. D.~M.,  1996, \mn@doi [\apj]
  {10.1086/177173}, \href
  {https://ui.adsabs.harvard.edu/abs/1996ApJ...462..563N} {462, 563}

\bibitem[\protect\citeauthoryear{{Purcell}, {Bullock}, {Tollerud}, {Rocha}  \&
  {Chakrabarti}}{{Purcell} et~al.}{2011}]{Purcell2011}
{Purcell} C.~W.,  {Bullock} J.~S.,  {Tollerud} E.~J.,  {Rocha} M.,
  {Chakrabarti} S.,  2011, \mn@doi [\nat] {10.1038/nature10417}, \href
  {https://ui.adsabs.harvard.edu/abs/2011Natur.477..301P} {477, 301}

\bibitem[\protect\citeauthoryear{{Quillen} et~al.,}{{Quillen}
  et~al.}{2018}]{Quillen2018}
{Quillen} A.~C.,  et~al., 2018, \mn@doi [\mnras] {10.1093/mnras/sty2077}, \href
  {https://ui.adsabs.harvard.edu/abs/2018MNRAS.480.3132Q} {480, 3132}

\bibitem[\protect\citeauthoryear{{Sch{\"o}nrich}, {Binney}  \&
  {Dehnen}}{{Sch{\"o}nrich} et~al.}{2010}]{Vsun}
{Sch{\"o}nrich} R.,  {Binney} J.,   {Dehnen} W.,  2010, \mn@doi [\mnras]
  {10.1111/j.1365-2966.2010.16253.x}, \href
  {https://ui.adsabs.harvard.edu/abs/2010MNRAS.403.1829S} {403, 1829}

\bibitem[\protect\citeauthoryear{{Schulz}, {Dehnen}, {Jungman}  \&
  {Tremaine}}{{Schulz} et~al.}{2013}]{Schulz13a}
{Schulz} A.~E.,  {Dehnen} W.,  {Jungman} G.,   {Tremaine} S.,  2013, \mn@doi
  [\mnras] {10.1093/mnras/stt073}, \href
  {https://ui.adsabs.harvard.edu/abs/2013MNRAS.431...49S} {431, 49}

\bibitem[\protect\citeauthoryear{{Ting}, {Rix}, {Bovy}  \& {van de Ven}}{{Ting}
  et~al.}{2013}]{Ting}
{Ting} Y.-S.,  {Rix} H.-W.,  {Bovy} J.,   {van de Ven} G.,  2013, \mn@doi
  [\mnras] {10.1093/mnras/stt1053}, \href
  {https://ui.adsabs.harvard.edu/abs/2013MNRAS.434..652T} {434, 652}

\bibitem[\protect\citeauthoryear{{Vasiliev} \& {Belokurov}}{{Vasiliev} \&
  {Belokurov}}{2020}]{SgrModel}
{Vasiliev} E.,  {Belokurov} V.,  2020, arXiv e-prints, \href
  {https://ui.adsabs.harvard.edu/abs/2020arXiv200602929V} {p. arXiv:2006.02929}

\bibitem[\protect\citeauthoryear{{Vasiliev}, {Belokurov}  \&
  {Erkal}}{{Vasiliev} et~al.}{2020}]{Vasiliev2020}
{Vasiliev} E.,  {Belokurov} V.,   {Erkal} D.,  2020, \mn@doi [\mnras]
  {10.1093/mnras/staa3673}, \href
  {https://ui.adsabs.harvard.edu/abs/2020MNRAS.tmp.3446V} {}

\bibitem[\protect\citeauthoryear{{Widmark} \& {Monari}}{{Widmark} \&
  {Monari}}{2019}]{p0-2}
{Widmark} A.,  {Monari} G.,  2019, \mn@doi [\mnras] {10.1093/mnras/sty2400},
  \href {https://ui.adsabs.harvard.edu/abs/2019MNRAS.482..262W} {482, 262}

\bibitem[\protect\citeauthoryear{{Widrow}, {Gardner}, {Yanny}, {Dodelson}  \&
  {Chen}}{{Widrow} et~al.}{2012}]{widrow12}
{Widrow} L.~M.,  {Gardner} S.,  {Yanny} B.,  {Dodelson} S.,   {Chen} H.-Y.,
  2012, \mn@doi [\apjl] {10.1088/2041-8205/750/2/L41}, \href
  {https://ui.adsabs.harvard.edu/abs/2012ApJ...750L..41W} {750, L41}

\bibitem[\protect\citeauthoryear{{Yanny} \& {Gardner}}{{Yanny} \&
  {Gardner}}{2013}]{yanny}
{Yanny} B.,  {Gardner} S.,  2013, \mn@doi [\apj] {10.1088/0004-637X/777/2/91},
  \href {https://ui.adsabs.harvard.edu/abs/2013ApJ...777...91Y} {777, 91}

\makeatother
\end{thebibliography}

\appendix

\section{\texttt{wendy}: A one-dimensional gravitational N-body code}\label{sec:wendy}

\texttt{wendy}\footnote{Available at \url{http://github.com/jobovy/wendy}} is a Python package that implements various methods for solving the one-dimensional $N$-body problem. $N$-body simulations in one dimension are much simpler than those in higher dimensions, because the gravitational force is constant with distance, rather than decreasing as, e.g., $1/r^2$ in three dimensions. The gravitational charge in one dimension has units of surface density, which means particles in one-dimensional simulation are also referred to as sheets and, in fact, the dynamics is equivalent to that of a set of parallel infinite sheets interacting through gravity in three dimensions. However, to keep to usual $N$-body nomenclature, we shall refer to them as particles here. 

For particles at positions $z_i$ with surface densities $\Sigma_i$, the gravitational force on particle $i$ from particle $j$ is
\begin{equation}
    F_{ij} = -2\pi G\,\Sigma_i\,\Sigma_j\,\mathrm{sign}(z_i-z_j)\,.
\end{equation}
The total force from all particles on a particle $i$ is therefore given by
\begin{equation}
    F_i = -2\pi G \Sigma_i \,\left[\Sigma(z_j > z_i) - \Sigma(z_j < z_i)] \right]\,,
\end{equation}
where $\Sigma(z_j > z_j)$ is the total surface density of particles with $z_j > z_i$ and $\Sigma(z_j < z_i)$ is the total surface density from particles at $z_j < z_i$. All of the forces in a one-dimensional $N$-body simulation can therefore be efficiently computed in $\mathcal{O}(N\log N)$ time using a simple sort of the positions.

To simulate the dynamics of a set of particles in one dimension, there are two options. The first option uses that because the gravitational force is constant with distance, it is constant on all particles as long as no particles cross; therefore, the dynamics can be exactly solved and all that is required is to resolve crossings of pairs of particles, which can also be done analytically (e.g., \citealt{Schulz13a}). However, this approach is slow (scaling as $\mathcal{O}(N^2\log N)$ to simulate a system for a dynamical time) and breaks down in the presence of an external force that is not harmonic. Therefore, we do not use it here, but \texttt{wendy} does implement this in a similar way as described by \citet{Schulz13a}.

The second option is to compute the force on each particle at a given time step exactly by sorting the particles, but to use a standard leapfrog integrator to advance the particles for a fixed time step, regardless of whether two particles cross in this time interval or not. In this case, it is straightforward to include any external force $F_\mathrm{ext}(z,t)$. This option is implemented in \texttt{wendy} as well, with different methods available for performing the sort (different implementations of quicksort, mergesort, and Python's timsort). This allows simulations with millions of particles to be run on a single core. Scaling to larger number of particles is difficult, because the limiting step is the sorting of the particles, which is difficult to parallelize.

\texttt{wendy} is implemented in Python with all actual $N$-body computations done in C. Arbitrary external forces can be used and are simply implemented in Python and automatically ported to C using \texttt{numba} \citep{numba}, which for simple forces can generated and compile C code that is as fast as native C code. \texttt{wendy} comes with a large set of examples, including the simulations of gravitational collapse and violent relaxation described by \citet{Schulz13a} and a simple model for the \emph{Gaia} phase-space spiral. \texttt{wendy} has an automated test suite, run upon every code push to GitHub and (currently) at least once per week on \texttt{Travis CI}, and has 100\,\% test line coverage. The code is available on the Python Package Index and can be installed with 
\begin{equation*}
    \texttt{pip install wendy}
\end{equation*}
or the latest development version can be downloaded from the code website and installed locally.

% Don't change these lines
\bsp	% typesetting comment
\label{lastpage}
\end{document}